\documentclass[12pt]{article}

\usepackage{epsf}
\usepackage{graphics}
\usepackage{epsfig}
\usepackage{amsmath}
\usepackage{amsfonts}
\usepackage{amssymb}

\begin{document}

\title{Dynamics of a mean spherical model with competing interactions}


\author{M O Hase and S R Salinas\\Instituto de F\'isica, Universidade de S\~ao Paulo \\ Caixa Postal 66318 05315-970, S\~ao Paulo, SP, Brazil\\ mhase@if.usp.br and ssalinas@if.usp.br}

\maketitle

\begin{abstract}
The Langevin dynamics of a $d$-dimensional mean spherical model with competing interactions along $m\leq d$ directions of a hypercubic lattice is analysed. After a quench at high temperatures, the dynamical behaviour is characterized by two distinct time scales associated with stationary and aging regimes. The asymptotic expressions for the autocorrelation and response functions, in supercritical, critical, and subcritical cases, were calculated. Aging effects, which are known to be present in the ferromagnetic version of this model system, are not affected by the introduction of competing interactions. 
\end{abstract}



\maketitle

\section{Introduction}

Non-equilibrium phenomena, as aging and violations of the fluctuation - dissipation theorem (FDT), have been attracting the attention of many investigators. A number of dynamical calculations for disordered as well as uniform magnetic model systems point out the occurrence of aging and violations of the FDT in a time evolution from a quench at high temperatures \cite{ZKH00, GL00, CST01}. Since there are no general principles to understand and classify these dynamical phenomena, it has been valuable to analyse the dynamical behavior of simple, analytically tractable, model systems. In the present work, a detailed investigation of the Langevin dynamics of an analytically tractable $d$-dimensional mean spherical model with competing interactions is reported.

Spin models with competing interactions are known to display a rich phase diagram, with multicritical points and modulated phases. In terms of the temperature $T$ and of a parameter $p$ gauging the strength of the competing interactions, the phase diagram of the axial next-nearest-neighbor Ising (or ANNNI\cite{S88, S92}) model displays a Lifshitz point, at the meeting of paramagnetic-ferromagnetic and paramagnetic-modulated critical lines, and an impressive sequence of modulated structures at low temperatures. The thermodynamic behavior of a spherical version of an Ising model with competing interactions has been originally analysed by Kalok and Obermair \cite{KO76}. A spherical analog of the ANNNI model, with the characterization of a Lifshitz point and the existence of ordered ferromagnetic and helical phases, has been introduced by Hornreich and coworkers\cite{HLS75}. The field behavior of this spherical analog of the ANNNI model has been investigated by Yokoi and coworkers\cite{YCS81}. With the exception of a few numerical works, as the analysis of an Ising model with both ferromagnetic and antiferromagnetic dipolar interactions in order to account for the behavior of ultrathin magnetic films\cite{dBMW00}, we are not aware of analytical investigations of the dynamics of statistical models with competing interactions.

In a recent paper, Godr\`{e}che and Luck \cite{GL00} reported a detailed analytical treatment of the Langevin dynamics of a $d$-dimensional ferromagnetic mean spherical model. The present work may be regarded as an extension of this analysis for a mean-spherical model with competing interactions. The particular results of Godr\`{e}che and Luck are recovered. 

The layout of this paper is as follows. The spherical model with competing interactions is introduced in Section 2. The Langevin dynamics, with the inclusion of a time-dependent Lagrange multiplier for implementing the spherical constraint, is analysed in Section 3, but the mathematical details of this analysis are left for the Appendix. In Section 4, some comments are made and the main conclusions are presented.

\section{Definition of the Model}

The grand canonical partition function,

\begin{eqnarray}
\Xi_{N}(\beta, \mu) = \int \exp \left[ -\beta H(\{S_{x}\}) - \beta \mu \sum_{x \in \Lambda_{N}} S_x^{2} \right] \prod_{x \in \Lambda_{N}} dS_{x},
\label{Z}
\end{eqnarray}

\noindent
subjected to a spherical constraint,

\begin{eqnarray}
\langle \sum_{x \in \Lambda_{N}} S_{x}^{2} \rangle = - \frac{1}{\beta} \frac{\partial}{\partial \mu} \ln \Xi_{N}(\beta, \mu) = N,
\label{constraint}
\end{eqnarray}

\noindent
is the trademark of a mean spherical model. The spin variables $S_{x}\in\mathbb{R}$ are continuous, $\beta$ is the (inverse) temperature, $\mu$ is a Lagrange multiplier that canonically ensures the spherical constraint, and $\Lambda_{N} = \{-L, -L+1, \cdots, L, L+1\}^{d}$ is a hypercubic lattice with $N$ sites. The Hamilton function is given by

\begin{eqnarray}
H(\{S_{x}\}) = - \frac{1}{2} \sum_{x, x^{\prime} \in \Lambda_{N}} J_{x, x^{\prime}} S_{x} S_{x^{\prime}},
\label{H}
\end{eqnarray}

\noindent
where

\begin{eqnarray}
J_{x, x^{\prime}} = \left\{
\begin{array}{lcll}
RJ & , & x - x^{\prime} = \pm e_{i}  & i \in \{1, \cdots, m\}   \\
SJ & , & x - x^{\prime} = \pm 2e_{i} & i \in \{1, \cdots, m\}   \\
J  & , & x - x^{\prime} = \pm e_{i}  & i \in \{m+1, \cdots, d\} \\
0  & , & \mbox{otherwise}            &
\end{array}
\right..
\label{J}
\end{eqnarray}

\noindent
This exchange integral describes the whole features of the model. There are nearest and next-nearest neighbor interactions along $m$ ($\leq d$) out of the $d$ directions of the hypercubic lattice; along the remaining $d-m$ directions, there are only ferromagnetic, $J>0$, nearest-neighbor interactions. We assume periodic boundary conditions along each direction ($S_{L+1}=S_{-L}$). Parameters $R$ and $S$ are free, but the scenario of competition takes place for $S<0$. In the particular (ferromagnetic) case analysed by Godr\`{e}che and Luck, $m=0$. The simple spherical analog of the ANNNI model is given by $m=1$ (with the parameter $p=-S/R$ gauging the strength of the competition).

The partition function can be obtained by standard procedures\cite{J72}. In the thermodynamic limit, the spherical constraint leads to the relation

\begin{eqnarray}
\beta(\mu) = \int \limits_{[- \pi, \pi]^{d}} \frac{d^{d}k}{2 (2\pi)^{d}} \frac{1}{\mu - \frac{1}{2} \hat J(k)}\,,
\label{beta}
\end{eqnarray}

\noindent
where $\beta$ is written in terms of the Lagrange multiplier $\mu$, and

\begin{eqnarray}
\mu \geq \mu_{c} := \frac{1}{2} \hat J(k_{c}) = \frac{1}{2} \sup_{k \in [-\pi,\pi]^{d}} \left\{ \hat J(k) \right\},
\label{muc}
\end{eqnarray}

\noindent
where

\begin{eqnarray}
\hat J(k) = 2J \bigg[R\sum_{i=1}^{m}\cos k_{i}+S\sum_{i=1}^{m}\cos(2k_{i})+\sum_{i=m+1}^{d}\cos k_{i}\bigg]
\label{hatJ}
\end{eqnarray}

\noindent
is the Fourier transform of the exchange integral. The critical wave vector $k_{c}$ comes from equation (\ref{muc}). Thus, one can write

\begin{eqnarray}
k_{c} = (\underbrace{q_{c}, \cdots, q_{c}}_{m}, \underbrace{0, \cdots, 0}_{d-m}),
\label{kc}
\end{eqnarray}

\noindent
where this point in the first Brillouin zone satisfies the condition (\ref{muc}), and is determined by the parameters $R$ and $S$. It is easy to see that

\begin{eqnarray}
q_{c} = \left\{
\begin{array}{lcl}
0       & , & R>0\quad\mbox{and}\quad S>-R/4 \\
 & & \\
\pi     & , & R<0\quad\mbox{and}\quad S>-R/4 \\
 & & \\
\pm\phi & , & S<-|R|/4
\end{array}
\right.,
\label{qc}
\end{eqnarray}

\noindent
where $\phi:=\arccos\left(-\frac{R}{4S}\right)$.

The sum rule (\ref{beta}), which defines the critical temperature $\beta (\mu _{c})$, also leads to the lower and upper critical dimensions, $d_{c}$ and $\overline{d}$, which are listed in Table 1, in terms of $d$, $m$, and the ratio $r=m/d$. Note that it is convenient to introduce and analyse five different cases. Also, note that these ingredients will be sufficient for characterizing the asymptotic dynamical behavior.

\begin{center}
\begin{tabular}{|c|c|c|l|l|} \hline
\rule[-2.0mm]{0mm}{6.5mm} & Case 1 & $|R|+4S \neq 0$ & $d_{c}=2$ & $\overline{d}=4$ \\
\cline{2-5}
\rule[-2.0mm]{0mm}{6.5mm} \raisebox{2.25ex}[0pt]{$m \neq 0$, $d-m \neq 0$} & Case 2 & $|R|+4S=0$ & $d_{c}=\frac{2}{2-r}$ & $\overline{d}=\frac{4}{2-r}$ \\
\hline
\rule[-2.0mm]{0mm}{6.5mm} & Case 3 & $|R|+4S \neq 0$ & $d_{c}=2$ & $\overline{d}=4$ \\ 
\cline{2-5}
\rule[-2.0mm]{0mm}{6.5mm} \raisebox{2.25ex}[0pt]{$m \neq 0$, $d-m=0$} & Case 4 & $|R|+4S=0$ & $d_{c}=4$ & $\overline{d}=8$ \\
\hline
\rule[-2.0mm]{0mm}{6.5mm} $m=0$, $d-m \neq 0$ & Case 5 & $\times$ & $d_{c}=2$ & $\overline{d}=4$ \\
\hline
\end{tabular}
\end{center}

\begin{center}
\textbf{Table 1:} Lower and upper critical dimensions.
\end{center}

\section{The Langevin dynamics}

The dynamics is assumed to be governed by the Langevin equation,

\begin{eqnarray}
\frac{\partial S_{x}(t)}{\partial t} = - \frac{\delta}{\delta S_{x}(t)} \left\{H[S_{x}](t) + \mu(t)\sum_{x \in\Lambda}S_{x}^{2}(t)\right\}+\xi_{x}(t),
\label{langevin}
\end{eqnarray}

\noindent
where $\{\xi_{x}(t)\}$ is a set of random variables such that

\begin{eqnarray}
\langle\xi_{x}(t)\rangle = 0 \qquad\mbox{and}\qquad \langle\xi_{x}(t)\xi_{x^{\prime}}(t^{\prime})\rangle = 2T\delta_{x, x^{\prime}}\delta(t-t^{\prime}).
\label{va}
\end{eqnarray}

\indent

In contrast to the static case, the Lagrange multiplier $\mu$ is now a funtion of time, ensuring the spherical constraint at each time $t$.

In this work, the calculations are limited to the long - time behaviour of autocorrelations and response functions. The analysis is restricted to the asymptotic expansions (for large $t^{\prime }$) of the autocorrelation,

\begin{eqnarray}
C(t,t^{\prime}) := \frac{1}{N}\sum_{x \in\Lambda_{N}}\langle S_{x}(t)S_{x}(t^{\prime}) \rangle = \frac{1}{N}\sum_{k \in\hat\Lambda_{N}}C_{k}(t,t^{\prime}),
\label{C}
\end{eqnarray}

\noindent
with $t>t^{\prime }$, where $C_{k}(t,t^{\prime })=\langle S_{k}(t)S_{-k}(t^{\prime })\rangle $ is a two-time correlation in the Fourier space $\hat{\Lambda}$, and the response function,

\begin{eqnarray}
R(t,t^{\prime}) := \frac{1}{N}\sum_{x \in\Lambda_{N}}\frac{\delta S_{x}(t)}{\delta h_{x}(t^{\prime})}\bigg|_{h\downarrow 0} = \frac{1}{N}\sum_{k \in\hat\Lambda_{N}}R_{k}(t,t^{\prime}),
\label{R}
\end{eqnarray}

\noindent
where $R_{k}(t,t^{\prime })=\delta \langle S_{x}(t)\rangle /\delta h_{k}(t^{\prime })$, and $h$ is just a small perturbation. According to the fluctuation-dissipation theorem, in a stationary regime these functions are related by the expression

\begin{eqnarray}
X(t, t^{\prime}) = \frac{T R(t, t^{\prime})}{\partial_{t^{\prime}} C(t, t^{\prime})} = 1.
\label{X}
\end{eqnarray}

\noindent
If $X(t, t^{\prime})\neq 1$ the theorem is violated, which suggests the introduction of an effective temperature $T/X(t, t^{\prime})$, larger than the heat-bath temperature $T$, and which is supposed to gauge a non-stationary behaviour of the system.

In order to calculate the two-time functions, one may first define the functional

\begin{eqnarray}
\psi[\mu](t) := \exp \left[ 4\int \limits_{0}^{t} \mu(t^{\prime}) dt^{\prime} \right],
\label{psi}
\end{eqnarray}

\noindent
which will be denoted by $\psi(t)$. By solving the differential equation (\ref{langevin}) and using the definition of $C_{k}(t, t^{\prime})$, one can show that 

\begin{eqnarray}
\nonumber C_{k}(t, t^{\prime}) & = & \frac{1}{\sqrt{\psi(t)\psi(t^{\prime})}}\bigg\{C_{k}(0, 0)\exp\left[\hat J(k)\left(t+t^{\prime}\right)\right] + \\
 & & + 2T\int\limits_{0}^{t^{\prime}}\exp\left[\hat J(k)\left(t+t^{\prime}-2t^{\prime\prime}\right)\right]\psi(t^{\prime\prime})dt^{\prime\prime}\bigg\},
\label{Cktt}
\end{eqnarray}

\noindent
where $C_{k}(0,0)$ is the initial condition. For a quench from a totally disordered state, at an effectively infinite temperature, one should take $C_{k}(0,0)=1$.

The autocorrelation is obtained from the spherical constraint $C(t,t)=1$ (see (\ref{C}) and (\ref{constraint})), that implies, in the thermodynamics limit, and for $t\geq 0$,

\begin{eqnarray}
\psi(t) = f(t) + 2T \int \limits_{0}^{t} f(t - t^{\prime}) \psi(t^{\prime}) dt^{\prime},
\label{volterra}
\end{eqnarray}

\noindent
with

\begin{eqnarray}
f(t) := \int\limits_{[-\pi,\pi]^{d}}\frac{d^{d}k}{\left(2\pi\right)^{d}}e^{2 \hat J(k)t} = \left[I_{0}(4Jt)\right]^{d-m}\left[\frac{1}{\pi}\int\limits_{0}^{\pi}e^{4Jtg(k)}dk\right]^{m},
\label{f}
\end{eqnarray}

\noindent
where $I_{0}(x)$ is the modified Bessel function of order zero, and

\begin{eqnarray}
g(k) := R\cos k+S\cos (2k)
\label{g}
\end{eqnarray}

\noindent
corresponds to the portion of the exchange integral responsible for the competition (if $S<0$).

The convolution product in equation (\ref{volterra}) suggests a solution by Laplace transform, which yields

\begin{eqnarray}
\psi(t) = \frac{1}{2\pi i}\int\limits_{\sigma - i\infty}^{\sigma + i\infty}dse^{st}\frac{\mathcal{L}[f](s)}{1-2T\mathcal{L}[f](s)},
\label{Lap}
\end{eqnarray}

\noindent
where $\sigma$ is larger than any poles of the integrand. The problem is now to determine $\psi$, since the autocorrelation,

\begin{eqnarray}
C(t, t^{\prime}) = \frac{1}{ \sqrt{ \psi(t) \psi(t^{\prime}) } } \left[ f \left( \frac{t + t^{\prime}}{2} \right) + 2T \int \limits_{0}^{t^{\prime}} dy f \left( \frac{t + t^{\prime}}{2} - y \right) \psi(y) \right],
\label{Cpsi}
\end{eqnarray}

\noindent
and the response function,

\begin{eqnarray}
R(t, t^{\prime}) = f \left( \frac{t - t^{\prime}}{2} \right) \sqrt{ \frac{\psi(t^{\prime})}{\psi(t)} },
\label{Rpsi}
\end{eqnarray}

\noindent
can be both written in terms of $\psi$.




The asymptotic behaviour of $f$ is

\begin{eqnarray}
f(t) \sim K_{p}\frac{e^{2\hat J(k_{c})t}}{t^{\gamma_{p}}},
\label{fasymp}
\end{eqnarray}

\noindent
where $p$ labels the set $p=\{R,S,m,d\}$. As it is fully discussed in the Appendix, the expressions for $K_{p}$ and $\gamma_{p}$, which are listed in Table 2, depend on the four parameters $\{R,S,m,d\}$.

\begin{center}
\begin{tabular}{|c|c|c|} \hline
\rule[-2.0mm]{0mm}{6.5mm}        & $K_{p}$ & $\gamma_{p}$ \\ \hline
\rule[-2.0mm]{0mm}{6.5mm} Case 1 & $2^{m}\left(8\pi J\right)^{-\frac{d}{2}}\left|g^{(2)}(q_{c})\right|^{-\frac{m}{2}}$ & $\frac{d}{2}$ \\ \hline
\rule[-2.0mm]{0mm}{6.5mm} Case 2 & $\left(8\pi J\right)^{-\frac{2d-m}{4}}\left(\frac{48}{\pi^{3}}\right)^{\frac{m}{4}}\Gamma\left(\frac{5}{4}\right)^{m}\left|g^{(4)}(q_{c})\right|^{-\frac{m}{4}}$ & $\frac{2d-m}{4}$ \\ \hline
\rule[-2.0mm]{0mm}{6.5mm} Case 3 & $2^{d}\left(8\pi J\right)^{-\frac{d}{2}}\left|g^{(2)}(q_{c})\right|^{-\frac{d}{2}}$ & $\frac{d}{2}$ \\ \hline
\rule[-2.0mm]{0mm}{6.5mm} Case 4 & $\left(8\pi J\right)^{-\frac{d}{4}}\left(\frac{48}{\pi^{3}}\right)^{\frac{d}{4}}\Gamma\left(\frac{5}{4}\right)^{d}\left|g^{(4)}(q_{c})\right|^{-\frac{d}{4}}$ & $\frac{d}{4}$ \\ \hline
\rule[-2.0mm]{0mm}{6.5mm} Case 5 & $\left(8\pi J\right)^{-\frac{d}{2}}$ & $\frac{d}{2}$ \\ \hline
\end{tabular}
\end{center}

\begin{center}
\textbf{Table 2:} $K_{p}$ and $\gamma_{p}$.
\end{center}

\indent

The behaviour of $\psi$ for large times demands the asymptotic expansion of $\mathcal{L}[f]$ for small values of $\epsilon:=s-2\hat J(k_{c})>0$. With the same notation, $r=m/d$, and $p=\{R,S,m,d\}$, one can calculate the asymptotic expression

\begin{eqnarray}
\mathcal{L}[f](s) \sim \left\{
\begin{array}{lcl}
\frac{F_{p} g_{p}}{\epsilon^{- \alpha_{p}}} & & 0 < d < d_{c} \\
 & & \\
F_{p} \left(- \ln \epsilon \right) & & d = d_{c} \\
 & & \\
A_{1} - F_{p}|g_{p}| \epsilon^{\alpha_{p}} & & d_{c} < d < \overline{d} \\
 & & \\
A_{1} - F_{p} \left(- \epsilon \ln \epsilon \right) & & d = \overline{d} \\
 & & \\
A_{1} - F_{p} \epsilon & & d > \overline{d}
\end{array}
\right.,
\label{Lf}
\end{eqnarray}

\noindent
with the coefficients given in Table 3. Note that $g^{(n)}$ is the $n$th derivative of $g$, given by equation (\ref{g}), which is different from $g_{p}$. Also, note that $\alpha _{p}=\gamma _{p}-1$.

\vspace{5mm}

\begin{center}
\begin{tabular}{|c|c|c|} \hline
 \rule[-2.0mm]{0mm}{6.5mm} Case & $F_{p}$ & $d$ \\ \hline\hline
 \rule[-2.0mm]{0mm}{6.5mm} Case 1 & $\times$ & $d<2$ \\ \cline{2-3}
 \rule[-2.0mm]{0mm}{6.5mm} & $2^{m}\left(8\pi J\right)^{-1}|g^{(2)}(q_{c})|^{-\frac{m}{2}}$ & $d=2$ \\ \cline{2-3}
 \rule[-2.0mm]{0mm}{6.5mm} \raisebox{2.25ex}[0pt]{$\alpha_{p}=\frac{d-2}{2}$} & $2^{m}\left(8\pi J\right)^{-\frac{d}{2}}|g^{(2)}(q_{c})|^{-\frac{m}{2}}$ & $2<d<4$ \\ \cline{2-3}
 \rule[-2.0mm]{0mm}{6.5mm} \raisebox{2.25ex}[0pt]{} & $2^{m}\left(8\pi J\right)^{-2}|g^{(2)}(q_{c})|^{-\frac{m}{2}}$ & $d=4$ \\ \cline{2-3}
 \rule[-2.0mm]{0mm}{6.5mm} \raisebox{2.25ex}[0pt]{$g_{p}=\Gamma\left(\frac{2-d}{2}\right)$} & $A_{2}$ & $d>4$ \\ \hline\hline
 \rule[-2.0mm]{0mm}{6.5mm} Case 2 & $\left(8\pi J\right)^{-\frac{2d-m}{4}}\left(\frac{48}{\pi^{3}}\right)^{\frac{m}{4}}\Gamma\left(\frac{5}{4}\right)^{m}|g^{(4)}(q_{c})|^{-\frac{m}{4}}$ & $d<\frac{4}{2-r}$ \\ \cline{2-3}
 \rule[-2.0mm]{0mm}{6.5mm} & $\left(8\pi J\right)^{-1}\left(\frac{48}{\pi^{3}}\right)^{\frac{m}{4}}\Gamma\left(\frac{5}{4}\right)^{m}|g^{(4)}(q_{c})|^{-\frac{m}{4}}$ & $d=\frac{4}{2-r}$ \\ \cline{2-3}
 \rule[-2.0mm]{0mm}{6.5mm} \raisebox{2.25ex}[0pt]{$\alpha_{p}=\frac{2d-m-4}{4}$} & $\left(8\pi J\right)^{-\frac{2d-m}{4}}\left(\frac{48}{\pi^{3}}\right)^{\frac{m}{4}}\Gamma\left(\frac{5}{4}\right)^{m}\left|g^{(4)}(q_{c})\right|^{-\frac{m}{4}}$ & $\frac{4}{2-r}<d<\frac{8}{2-r}$ \\ \cline{2-3}
 \rule[-2.0mm]{0mm}{6.5mm} \raisebox{2.25ex}[0pt]{} & $\left(8\pi J\right)^{-2}\left(\frac{48}{\pi^{3}}\right)^{\frac{m}{4}}\Gamma\left(\frac{5}{4}\right)^{m}|g^{(4)}(q_{c})|^{-\frac{m}{4}}$ & $d=\frac{8}{2-r}$ \\ \cline{2-3}
 \rule[-2.0mm]{0mm}{6.5mm} \raisebox{2.25ex}[0pt]{$g_{p}=\Gamma\left(\frac{4-2d+m}{4}\right)$} & $A_{2}$ & $d>\frac{8}{2-r}$ \\ \hline\hline
 \rule[-2.0mm]{0mm}{6.5mm} Case 3 & $2^{d}\left(8\pi J\right)^{-\frac{d}{2}}|g^{(2)}(q_{c})|^{-\frac{d}{2}}$ & $d<2$ \\ \cline{2-3}
 \rule[-2.0mm]{0mm}{6.5mm} & $2^{d}\left(8\pi J\right)^{-1}|g^{(2)}(q_{c})|^{- 1}$ & $d=2$ \\ \cline{2-3}
 \rule[-2.0mm]{0mm}{6.5mm} \raisebox{2.25ex}[0pt]{$\alpha_{p}=\frac{d-2}{2}$} & $2^{d}\left(8\pi J\right)^{-\frac{d}{2}}|g^{(2)}(q_{c})|^{-\frac{d}{2}}$ & $2<d<4$ \\ \cline{2-3}
 \rule[-2.0mm]{0mm}{6.5mm} \raisebox{2.25ex}[0pt]{} & $2^{d}\left(8\pi J\right)^{-2}|g^{(2)}(q_{c})|^{-2}$ & $d=4$ \\ \cline{2-3}
 \rule[-2.0mm]{0mm}{6.5mm} \raisebox{2.25ex}[0pt]{$g_{p}=\Gamma\left(\frac{2-d}{2}\right)$} & $A_{2}$ & $d>4$ \\ \hline\hline
 \rule[-2.0mm]{0mm}{6.5mm} Case 4 & $\left(8\pi J\right)^{-\frac{d}{4}}\left(\frac{48}{\pi^{3}}\right)^{\frac{d}{4}}\Gamma\left(\frac{5}{4}\right)^{d}|g^{(4)}(q_{c})|^{-\frac{d}{4}}$ & $d<4$ \\ \cline{2-3}
 \rule[-2.0mm]{0mm}{6.5mm} & $\left(8\pi J\right)^{-1}\left(\frac{48}{\pi^{3}}\right)\Gamma\left(\frac{5}{4}\right)^{4}|g^{(4)}(q_{c})|^{- 1}$ & $d=4$ \\ \cline{2-3}
 \rule[-2.0mm]{0mm}{6.5mm} \raisebox{2.25ex}[0pt]{$\alpha_{p}=\frac{d-4}{4}$} & $\left(8\pi J\right)^{-\frac{d}{4}}\left(\frac{48}{\pi^{3}}\right)^{\frac{d}{4}}\Gamma\left(\frac{5}{4}\right)^{d}|g^{(4)}(q_{c})|^{-\frac{d}{4}}$ & $4<d<8$ \\ \cline{2-3}
 \rule[-2.0mm]{0mm}{6.5mm} \raisebox{2.25ex}[0pt]{} & $\left(8\pi J\right)^{-2}\left(\frac{48}{\pi^{3}}\right)^{2}\Gamma\left(\frac{5}{4}\right)^{8}|g^{(4)}(q_{c})|^{-2}$ & $d=8$ \\ \cline{2-3}
 \rule[-2.0mm]{0mm}{6.5mm} \raisebox{2.25ex}[0pt]{$g_{p}=\Gamma\left(\frac{4-d}{4}\right)$} & $A_{2}$ & $d>8$ \\ \hline\hline
 \rule[-2.0mm]{0mm}{6.5mm} Case 5 & $\left(8\pi J\right)^{-\frac{d}{2}}$ & $d<2$ \\ \cline{2-3}
 \rule[-2.0mm]{0mm}{6.5mm} & $\left(8\pi J\right)^{-1}$ & $d=2$ \\ \cline{2-3}
 \rule[-2.0mm]{0mm}{6.5mm} \raisebox{2.25ex}[0pt]{$\alpha_{p}=\frac{d-2}{2}$} & $\left(8\pi J\right)^{-\frac{d}{2}}$ & $2<d<4$ \\ \cline{2-3}
 \rule[-2.0mm]{0mm}{6.5mm} & $\left(8\pi J\right)^{-2}$ & $d=4$ \\ \cline{2-3}
 \rule[-2.0mm]{0mm}{6.5mm} \raisebox{2.25ex}[0pt]{$g_{p}=\Gamma\left(\frac{2-d}{2}\right)$} & $A_{2}$ & $d>4$ \\ \hline
\end{tabular}
\end{center}

\begin{center}
\textbf{Table 3:} $F_{p}$, $g_{p}$, and $\alpha_{p}$.
\end{center}

\vspace{5mm}

\indent

The next step is the determination of the asymptotic behaviour of $\psi$. The calculations are separated in three parts, each of them corresponding to a different temperature regime.

\subsection{Supercritical dynamics}

If a system is quenched from a highly disordered state (for instance, the system may have an effectively infinite temperature) to $T>T_{c}$, the function $\psi$ has an asymptotic exponential behaviour

\begin{eqnarray}
\psi(t) \sim e^{t/\tau_{p}}\,,
\label{psi>}
\end{eqnarray}

\noindent
where $\tau_{p}$ is related to the characteristic time. This behaviour indicates the decay of the system to an equilibrium state in finite time. In this situation the autocorrelation,

\begin{eqnarray}
C(t, t^{\prime}) \sim C(\tau) = T \int \limits_{\tau}^{\infty} dy f \left( \frac{y}{2} \right) e^{- \frac{y}{2 \tau_{p}}},
\label{C>}
\end{eqnarray}

\noindent
and the response function,

\begin{eqnarray}
R(t, t^{\prime}) \sim R(\tau) = f \left( \frac{\tau}{2} \right) e^{- \frac{\tau}{2 \tau_{p}} },
\label{R>}
\end{eqnarray}

\noindent
depend on the time difference $\tau$ only, and the fluctuation - dissipation theorem is satisfied,

\begin{eqnarray}
X(t, t^{\prime}) \sim 1.
\label{X>}
\end{eqnarray}

\subsection{Critical dynamics}

In contrast to the other cases, the critical dynamical behaviour depends on the dimension of the system. In the following calculations, it will always be assumed that $d$ is larger than the lower critical dimension $d_{c}$, so that $T_{c}\neq 0$; in other words, the occurrence of a phase transition is assumed.

The asymptotic behaviour of $\psi$ is given by

\begin{eqnarray}
\psi(t) \sim \left\{
\begin{array}{lcl}
\frac{e^{2 \hat J(k_{c})t}}{t^{1-\alpha_{p}}} & , & d_{c}<d<\overline{d} \\
 & & \\
\frac{e^{2 \hat J(k_{c})t}}{\ln t}            & , & d=\overline{d} \\
 & & \\
e^{2 \hat J(k_{c})t}                          & , & d>\overline{d}
\end{array}
\right.\,,
\label{psi=}
\end{eqnarray}

\noindent
which is sensitive to the dimension.

Two time scales arise in the analysis of the critical dynamics:

\medskip

\noindent
(i) In the stationary regime, $1\sim\tau\ll t^{\prime}$, both the autocorrelation,

\begin{eqnarray}
C(t, t^{\prime}) \sim C_{eq, c}(\tau),
\label{C=st}
\end{eqnarray}

\noindent
with

\begin{eqnarray}
C_{eq, c}(\tau) := T_{c} \int \limits_{\tau}^{\infty} dy f \left( \frac{y}{2} \right) e^{- \hat J(k_{c})y},
\label{Ceqc}
\end{eqnarray}

\noindent
and the response function,

\begin{eqnarray}
R(t, t^{\prime}) \sim f \left( \frac{\tau}{2} \right) e^{- \hat J(k_{c}) \tau},
\label{R=st}
\end{eqnarray}

\noindent
are invariant under time translation. The fluctuation - dissipation theorem is satisfied with $X(t,t^{\prime})\sim 1$. The choice $\tau\sim 1$ precludes the system to decay from the stationary state, which suggests the occurrence of aging for larger values of $\tau$.

\medskip

\noindent
(ii) For $1\ll\tau\sim t^{\prime}$, it is convenient to define

\begin{eqnarray}
x := \frac{t}{t^{\prime}}.
\label{x}
\end{eqnarray}

\noindent
In this regime, the autocorrelation,

\begin{eqnarray}
C(t,t^{\prime})\sim\left\{
\begin{array}{lcl}
\frac{2K_{p}T_{c}2^{\gamma_{p}}}{\gamma_{p}-1}t^{\prime^{1-\gamma_{p}}}\frac{x^{1-\frac{\gamma_{p}}{2}}\left(x-1\right)^{1-\gamma_{p}}}{x+1} & , & d_{c}<d<\overline{d} \\
 & & \\
\frac{T_{c}K_{p}2^{\gamma_{p}}}{\gamma_{p}-1}t^{\prime^{1-\gamma_{p}}}\sqrt{1+\frac{\ln x}{\ln t^{\prime}}} \times & & \\
 \times \left[\left(x-1\right)^{1-\gamma_{p}}-\left(x+1\right)^{1-\gamma_{p}}\right] & , & d=\overline{d} \\
 & & \\
\frac{2^{\gamma_{p}}K_{p}T_{c}}{\gamma_{p}-1}t^{\prime^{1-\gamma_{p}}} \left[ \left(x-1\right)^{1-\gamma_{p}} - \left(x+1\right)^{1-\gamma_{p}} \right] & , & d>\overline{d}
\end{array}
\right.,
\label{C=ag}
\end{eqnarray}

\noindent
and the response function,

\begin{eqnarray}
R(t,t^{\prime})\sim\left\{
\begin{array}{lcl}
2^{\gamma_{p}}K_{p}t^{\prime^{-\gamma_{p}}}x^{\frac{1-\alpha_{p}}{2}}\left(x-1\right)^{-\gamma_{p}} & , & d_{c}<d<\overline{d} \\
 & & \\
2^{\gamma_{p}}K_{p}t^{\prime^{-\gamma_{p}}}\left(x-1\right)^{-\gamma_{p}}\sqrt{1+\frac{\ln x}{\ln t^{\prime}}} & , & d=\overline{d} \\
 & & \\
2^{\gamma_{p}}K_{p}t^{\prime^{-\gamma_{p}}}\left(x-1\right)^{-\gamma_{p}} & , & d>\overline{d}
\end{array}
\right.\,,
\label{R=ag}
\end{eqnarray}

\noindent
show that the time translation invariance is broken, and aging effects are observed.


The asymptotic behaviour of the fluctuation-dissipation ratio is calculated from the equations (\ref{C=ag}) and (\ref{R=ag}), which yield

\begin{eqnarray}
X(t,t^{\prime})\sim\left\{
\begin{array}{lcl}
\frac{\left(\gamma_{p}-1\right)\left(x+1\right)^{2}}{\left(\gamma_{p}x+\gamma_{p}-2\right)\left(x+1\right)-2\left(x-1\right)} & , & d_{c}<d<\overline{d} \\
 & & \\
\frac{2\left(\gamma_{p}-1\right)\ln t^{\prime}}{ 2\left(\gamma_{p}-1\right)\left[1+\left(\frac{x-1}{x+1}\right)^{\gamma_{p}}\right]\ln t^{\prime} - \left(x-1\right)\left[1-\left(\frac{x-1}{x+1}\right)^{\gamma_{p}-1}\right] }
& , & d=\overline{d} \\
 & & \\
\frac{1}{1+\left(\frac{x-1}{x+1}\right)^{\gamma_{p}}} & , & d>\overline{d}
\end{array}
\right.\,.
\label{X=ag}
\end{eqnarray}

\noindent
Note that $x\sim 1$ is the stationary limit, and $X(t,t^{\prime})\sim 1$ in this case.

\subsection{Subcritical Dynamics}

Again, the occurrence of a phase transition is required, and the calculations are performed for $d>d_{c}$. The asymptotic behaviour of $\psi$ is given by

\begin{eqnarray}
\psi(t) \sim \frac{f(t)}{M_{eq}^{4}}\,,
\label{psi<}
\end{eqnarray}

\noindent
where

\begin{eqnarray}
M_{eq}^{2} := 1 - \frac{T}{T_{c}}
\label{Meq}
\end{eqnarray}

\noindent
is the square of the static magnetisation.

In the stationary regime, $1\sim\tau\ll t^{\prime}$, the autocorrelation,

\begin{eqnarray}
C(t,t^{\prime}) \sim M_{eq}^{2}+\left(1-M_{eq}^{2}\right)C_{eq,c}(\tau),
\label{C<st}
\end{eqnarray}

\noindent
and the response function,

\begin{eqnarray}
R(t,t^{\prime}) \sim f\left(\frac{\tau}{2}\right)e^{-\hat J(k_{c})\tau},
\label{R<st}
\end{eqnarray}

\noindent
depend on $\tau$, and the fluctuation - dissipation theorem is asymptotically satisfied.

In the aging time scale, $1\ll\tau\sim t^{\prime}$, the autocorrelation,

\begin{eqnarray}
C(t, t^{\prime}) \sim M_{eq}^{2} \left[ \frac{4x}{ \left( x + 1 \right)^{2} } \right]^{ \frac{\gamma_{p}}{2} },
\label{C<ag}
\end{eqnarray}

\noindent
and the response function,

\begin{eqnarray}
R(t, t^{\prime}) \sim K_{p} 2^{\gamma_{p}} t^{\prime^{- \gamma_{p}}} x^{ \frac{\gamma_{p}}{2} } \left( x - 1 \right)^{- \gamma_{p}},
\label{R<ag}
\end{eqnarray}

\noindent
are not invariant under time translation. One may calculate

\begin{eqnarray}
\lim_{\tau \rightarrow \infty} \lim_{t^{\prime} \rightarrow \infty} C(t, t^{\prime}) = M_{eq}^{2} = 1 - \frac{T}{T_{c}},
\label{plateau}
\end{eqnarray}

\noindent
which is the analogous of Edwards-Anderson order parameter. This is a connection between the two time scales, and one can also interpolate the autocorrelation as

\begin{eqnarray}
C(t, t^{\prime}) \sim \left( 1 - M_{eq}^{2} \right) C_{eq, c}(\tau) + M_{eq}^{2} \left[ \frac{4x}{\left( x + 1 \right)^{2}} \right]^{\frac{\gamma_{p}}{2}}.
\label{C<stag}
\end{eqnarray}

\noindent
The fluctuation-dissipation ratio is

\begin{eqnarray}
X(t, t^{\prime}) \sim \frac{2TK_{p}}{\gamma_{p}M_{eq}^{2}}t^{\prime^{1-\gamma_{p}}}\left(\frac{x+1}{x-1}\right)^{1+\gamma_{p}}.
\label{X<ag}
\end{eqnarray}

\section{Conclusions}

In this work, the Langevin dynamics of a $d$-dimensional mean spherical model on a hypercubic lattice with nearest-neighbor ($J$ and $RJ$) interactions and the addition of extra next-nearest-neighbor ($SJ$) interactions along $m\leq d$ directions was analysed. For $S<0$ there is a scenario of competition between ferromagnetic and antiferromagnetic interactions, with the occurrence of an ordered modulated region in the phase diagram (in terms of the temperature $T$ of the heat bath and the competition parameter $p=-S/R$). The asymptotic expressions (for large values of time $t^{\prime}$) for the autocorrelation, $C\left( t,t^{\prime }\right) $, and the response function, $R\left( t,t^{\prime }\right) $, with $t>t^{\prime }$, were obtained, and the validity of the fluctuation-dissipation ratio, $X\left( t,t^{\prime }\right)=TR\left( t,t^{\prime }\right) /\partial _{t^{\prime }}C\left( t,t^{\prime}\right)$, was checked.

The addition of competing interactions does not change the qualitative dynamical behaviour as compared to the ferromagnetic case (case 5 in this work), which has been analysed in detail by Godr\`{e}che and Luck\cite{GL00}. The supercritical dynamics is trivial. The asymptotic forms of the two-time functions are translational invariant, $X\left( t,t^{\prime }\right) \sim 1$, and the system reaches equilibrium in a finite time. In the critical and subcritical cases, one is led to consider two distinct natural time scales: (i) For $1\sim \tau \ll t^{\prime }$, the two-time functions depend on the difference $\tau =t-t^{\prime }$ only, and the fluctuation-dissipation theorem holds; (ii) If $1\ll \tau \sim t^{\prime }$, in general both the autocorrelation and the response function\footnote{Except the response function in critical dynamics for $d>\overline{d}$.} depend on $t$ and $t^{\prime }$ (instead of depending on $\tau $ only). This lack of translational invariance leads to violations of the fluctuation-dissipation theorem, and to a system that ages with time. 

\section{Acknowledgements}

The authors are indebted to D. A. Stariolo and D. H. U. Marchetti for many comments, and are thankful to A. Mauger for carefully reading the manuscript. The authors thank J. M. Luck for very useful observations. This work had the financial support of the Brazilian agencies FAPESP, CAPES, and CNPq.

\section*{Appendix}

The simple ferromagnetic model (case 5 in the classification of this work) is used in order to give a detailed account of the calculations. Results for the other cases can be obtained by analogous manipulations.

\subsection*{Lower critical dimension}

The lower critical dimension, $d_{c}$, is established by the spherical constraint (\ref{beta}) at the critical value $\mu=\mu_{c}$,

\begin{eqnarray}
\nonumber \hat J(k_{c}) - \hat J(k) & = & J \left[c_{2}\sum_{i=1}^{m}\left(k_{i}-q_{c}\right)^{2} + \sum_{i=m+1}^{d}k_{i}^{2}\right] - \frac{Jc_{3}}{3}\sum_{i=1}^{m}\left(k_{i}-q_{c}\right)^{3}+ \\
 & & +\frac{J}{12}\left[c_{4}\sum_{i=1}^{m}\left(k_{i}-q_{c}\right)^{4}+\sum_{i=m+1}^{d}k_{i}^{4}\right]+\cdots,
\label{a1}
\end{eqnarray}

\noindent
where $c_{2}:=R\cos q_{c}+4S\cos(2q_{c})$, $c_{3}:=R\sin q_{c}+8S\sin(2q_{c})$, and $c_{4}:=-R\cos q_{c}-16S\cos(2q_{c})$. It is easy to see that $c_{2}\geq 0$, and $c_{2}=0$ if and only if $R+4S=0$ (corresponding to $q_{c}=0$) and $R-4S=0$ (corresponding to $q_{c}=\pi$). In these cases $c_{3}$ also vanishes and therefore the fourth-order term becomes relevant to characterize the critical behaviour.

For the fifth case, the (inverse) critical temperature is

\begin{eqnarray}
\nonumber \beta(\mu_{c}) & = & \frac{1}{J} \int\limits_{B_{\delta}} \frac{d^{d}k}{\left(2\pi\right)^{d}} \frac{1}{\sum_{i=1}^{d}k_{i}^{2}+\mathcal{O}(\delta^{3})} + \int\limits_{\hat\Lambda\setminus B_{\delta}} \frac{d^{d}k}{\left(2\pi\right)^{d}} \frac{1}{\hat J(k_{c})-\hat J(k)} \\
\nonumber & = & \frac{1}{J} \int\limits_{B_{\delta}} \frac{d^{d}k}{\left(2\pi\right)^{d}} \frac{1}{\sum_{i=1}^{d}k_{i}^{2}} + \mathcal{O}(\delta) \\
 & = & \frac{1}{J} \frac{2\pi^{\frac{d}{2}}}{\Gamma \left(\frac{d}{2}\right)} \int\limits_{0}^{\delta} \frac{dk k^{d-3}}{\left(2\pi\right)^{d}} + \mathcal{O}(\delta),
\label{a2}
\end{eqnarray}

\noindent
where $B_{\delta}$ is an open ball of radius $\delta$ centered at $(0,\cdots,0)$, and in the last step the (hyper) spherical coordinates were invoked. The integral converges for $d>2$, establishing $d_{c}=2$. In this work, $h=\mathcal{O}(x)$ means that $h$ is of order $x$ or less than it; by $h=o(x)$, it means that $h$ is of order less than $x$.

\subsection*{Initial conditions}

From equation (\ref{C}), the autocorrelation in Fourier space at $t=t^{\prime}=0$ is given by

\begin{eqnarray}
C_{k}(0, 0) = \frac{1}{N} \sum_{x, x^{\prime} \in \Lambda_{N}} \langle S_{x}(0) S_{x^{\prime}}(0)\rangle e^{ik(x-x^{\prime})},
\label{a3}
\end{eqnarray}

\noindent
where

\begin{eqnarray}
\langle S_{x}(0) S_{x^{\prime}}(0)\rangle = \left\{
\begin{array}{lcl}
\langle S_{x}(0)\rangle \langle S_{x^{\prime}}(0)\rangle & , & x \neq x^{\prime} \\
\langle S_{x}^{2}(0)\rangle                              & , & x = x^{\prime}
\end{array}
\right.
\label{a4}
\end{eqnarray}

\noindent
for an ``infinite temperature'' condition. In this highly disordered situation, from the spherical constraint $C(t,t)=1$ at $t=0$, one has

\begin{eqnarray}
\nonumber N & = & \sum_{x \in \Lambda_{N}} \langle S_{x}^{2}(0)\rangle \\
            & = & N \langle S_{x}^{2}(0)\rangle,
\label{a5}
\end{eqnarray}

\noindent
from which $\langle S_{x}^{2}(0)\rangle=1$.

Therefore, $\langle S_{x}(0) S_{x^{\prime}}(0)\rangle=\delta_{x,x^{\prime}}$, which yields $C_{k}(0,0)=1$.

\subsection*{Asymptotic behaviour of $f$}

For large $t$, and choosing $\delta \ll 1$ such that $t^{-1/2}\ll\delta\ll t^{-1/4}$, one shows that (case 5)

\begin{eqnarray}
\nonumber f(t) & = & \left[ \frac{1}{\pi} \int \limits_{0}^{\pi} dk e^{4Jt \cos k} \right]^{d} \\
\nonumber & = & \left[ \frac{1}{\pi} \int \limits_{0}^{\delta} dk e^{4Jt \left(1 - \frac{k^{2}}{2} + \mathcal{O}(\delta^{4}) \right)} + \frac{1}{\pi} \int \limits_{\delta}^{\pi} dk e^{4Jt \cos k} \right]^{d} \\
\nonumber & = & \left[ \frac{e^{4Jt}}{\pi} \int \limits_{0}^{\delta} dk e^{- 2Jt k^{2}} \left( 1 + \mathcal{O}(t\delta^{4}) \right) + \frac{1}{\pi} \int \limits_{\delta}^{\pi} dk e^{4Jt \cos k} \right]^{d} \\
\nonumber & = & \left[ \frac{e^{4Jt}}{\pi} \frac{1}{\sqrt{2Jt}} \frac{\sqrt{\pi}}{2} \left( \mbox{erf}( \sqrt{2Jt}\delta) + \mathcal{O}(t^{3/2} \delta^{4}) \right) + \mathcal{O}(e^{4Jt \cos \delta}) \right]^{d} \\
 & \sim & \frac{e^{4Jdt}}{\left(8 \pi Jt\right)^{\frac{d}{2}}}.
\label{a6}
\end{eqnarray}

\indent

In general, one has equation (\ref{fasymp}) with $K_{p}$ and $\gamma_{p}$ given in Table 2.

\subsection*{Asymptotic behaviour of $\mathcal{L}[f](s)$}

One should consider

\begin{eqnarray}
\mathcal{L}[f](s) = \int\limits_{[- \pi, \pi]^{d}} \frac{d^{d}k}{\left(2\pi\right)^{d}} \frac{1}{\epsilon + 2\hat J(k_{c}) - 2\hat J(k)},
\label{a7}
\end{eqnarray}

\noindent
where $\epsilon:=s-2\hat J(k_{c})>0$, and recall that $k_{c}=0$ and $2\hat J(k)=4J\sum_{i=1}^{d}\cos k_{i}$ in case 5. In order to obtain the Laplace transform of $f$ in the vicinity of $2\hat J(k_{c})$ (or $\epsilon\sim 0$), and using the same notation as  in equation (\ref{a2}), this expression is rewritten in the form

\begin{eqnarray}
\nonumber \mathcal{L}[f](s) & = & \int\limits_{B_{\delta}} \frac{d^{d}k}{\left(2\pi\right)^{d}} \frac{1}{\epsilon + 2\hat J(k_{c}) - \left[ 2\hat J(k_{c}) - 2J \sum_{i=1}^{d}k_{i}^{2} + \mathcal{O}(\delta^{4})\right]} + \\
 & & + \int\limits_{[-\pi, \pi]^{d}\setminus B_{\delta}} \frac{d^{d}k}{\left(2\pi\right)^{d}} \frac{1}{\epsilon + 2\hat J(k_{c}) - 2\hat J(k)}.
\label{a8}
\end{eqnarray}

\indent

Changing the first term of (\ref{a8}) to (hyper) spherical coordinates, and since the second term is analytic in $\epsilon$, it is possible to write

\begin{eqnarray}
\mathcal{L}[f](s) = \frac{\epsilon^{\frac{d-2}{2}}}{\left(8\pi J\right)^{\frac{d}{2}} \Gamma\left(\frac{d}{2}\right)} \int\limits_{0}^{\frac{2J\delta^{2}}{\epsilon}} \frac{dkk^{\frac{d}{2}-1}}{k+1} + \mathcal{O}(\delta^{2}) + \sum_{j=1}^{\infty} \left(-1\right)^{j-1} A_{j}^{\prime} \epsilon^{j-1},
\label{a9}
\end{eqnarray}

\noindent
where

\begin{eqnarray}
A_{j}^{\prime} := \int\limits_{[-\pi, \pi]^{d}\setminus B_{\delta}} \frac{d^{d}k}{\left(2\pi\right)^{d}} \frac{1}{\left[ 2\hat J(k_{c}) - 2\hat J(k)\right]^{j}}.
\label{a10}
\end{eqnarray}

\indent

Also, define

\begin{eqnarray}
A_{j} := \int\limits_{[-\pi, \pi]^{d}} \frac{d^{d}k}{\left(2\pi\right)^{d}} \frac{1}{\left[ 2\hat J(k_{c}) - 2\hat J(k)\right]^{j}}.
\label{a11}
\end{eqnarray}

\noindent
From equation (\ref{beta}), it is easily seen that $A_{1}=\frac{1}{2T_{c}}$.

Let $d=2q$. For even dimensions, and choosing $\epsilon\ll\delta^{2}\ll 1$ such that $\epsilon\ln\delta\ll\delta^{2}$ and $\delta^{2}\ll|\epsilon\ln\epsilon|$ (a possible choice is given by $\delta^{2}=\epsilon\sqrt{|\ln\epsilon\,\ln\delta|}$), the integral in equation (\ref{a9}) is

\begin{eqnarray}
\nonumber \epsilon^{q-1}\int\limits_{0}^{\frac{2J\delta^{2}}{\epsilon}} \frac{dkk^{q-1}}{k+1} & = & \epsilon^{q-1} \int\limits_{1}^{1+\frac{2J\delta^{2}}{\epsilon}} \frac{du}{u}\sum_{m=0}^{q-1}{q-1 \choose m}\left(-1\right)^{q-1-m}u^{m} \\
\nonumber & = & \left\{
\begin{array}{lcl}
-\ln\epsilon + \mathcal{O}(\ln\delta^{2})                                              & , & q=1 \quad (d=2) \\
 & & \\
\epsilon\ln\epsilon + \mathcal{O}(\delta^{2})                                          & , & q=2 \quad (d=4) \\
 & & \\
\left(-1\right)^{q}\epsilon^{q-1}\ln\epsilon + \mathcal{O}\left(\delta^{2(q-1)}\right) & , & q>2 \quad (d>4)
\end{array}
\right.. \\
\label{a12}
\end{eqnarray}

\indent

On the other hand, for non-even dimensions $d\in(0,2)$, one sees that

\begin{eqnarray}
\nonumber \int\limits_{0}^{\frac{2J\delta^{2}}{\epsilon}} \frac{dkk^{\frac{d}{2}-1}}{k+1} & = & \Gamma\left(\frac{d}{2}\right)\Gamma\left(1-\frac{d}{2}\right) - \int\limits_{\frac{2J\delta^{2}}{\epsilon}}^{\infty}\frac{dkk^{\frac{d}{2}-1}}{k+1} \\
 & = & \Gamma\left(\frac{d}{2}\right)\Gamma\left(1-\frac{d}{2}\right) + \mathcal{O}\left( \left[\frac{\epsilon}{\delta^{2}}\right]^{\left|1-d/2\right|} \right).
\label{a13}
\end{eqnarray}

\indent

An analytic continuation from $(0,2)$ to $\mathbb{R}\setminus\mathbb{Z}$ using the functional equation (\ref{a13}) leads to

\begin{eqnarray}
\int\limits_{0}^{\frac{2J\delta^{2}}{\epsilon}} \frac{dkk^{\frac{d}{2}-1}}{k+1} \sim \Gamma\left(\frac{d}{2}\right)\Gamma\left(1-\frac{d}{2}\right), \qquad d\in\mathbb{R}\setminus\mathbb{Z}.
\label{a14}
\end{eqnarray}

\indent

Therefore,

\begin{eqnarray}
\nonumber \mathcal{L}[f](s) & = & \sum_{j=1}^{\infty} \left(-1\right)^{j-1} A_{j}^{\prime} \epsilon^{j-1} + \mathcal{O}(\delta^{2}) + \\
\nonumber & & + \left\{
\begin{array}{lcl}
\left(8\pi J\right)^{- \frac{d}{2}} \Gamma\left(1-\frac{d}{2}\right)\epsilon^{- \frac{2-d}{2}} + \mathcal{O}(\delta^{-(2-d)}) & , & d<2 \\
 & & \\
\left(8\pi J\right)^{-1} \left(-\ln\epsilon\right) + \mathcal{O}(\ln\delta^{2}) & , & d=2 \\
 & & \\
\left(8\pi J\right)^{- \frac{d}{2}} \Gamma\left(1-\frac{d}{2}\right)\epsilon^{\frac{d-2}{2}} + \mathcal{O}(\delta^{d-2}) & , & 2<d<4 \\
 & & \\
\left(8\pi J\right)^{-2} \epsilon\ln\epsilon + \mathcal{O}(\delta^{2}) & , & d=4 \\
 & & \\
\mathcal{O}(\delta^{2}) & , & d>4
\end{array}
\right.. \\
\label{a15}
\end{eqnarray}

\indent

One should now analyse the behaviour of $\sum_{j=1}^{\infty} \left(-1\right)^{j-1} A_{j}^{\prime} \epsilon^{j-1}$ for $\epsilon\sim 0$. First, note that $A_{j}$ (equation (\ref{a11})) is finite for $d>2j$. In this case

\begin{eqnarray}
\nonumber A_{j}^{\prime} & = & A_{j} - \int\limits_{B_{\delta}} \frac{d^{d}k}{\left(2\pi\right)^{d}} \frac{1}{\left[2\hat J(k_{c})-2\hat J(k)\right]^{j}} \\
\nonumber & = & A_{j} - \frac{2\pi^{\frac{d}{2}}}{\left(2\pi\right)^{d}\Gamma\left(\frac{d}{2}\right)\left(2J\right)^{j}} \int\limits_{0}^{\delta} dk k^{d-1-2j} + \mathcal{O}(\delta^{2}) \\
 & = & A_{j} + \mathcal{O}(\delta^{d-2j}) + \mathcal{O}(\delta^{2}).
\label{a16}
\end{eqnarray}

\indent

For $d\leq 2j$ the integral $A_{j}$ diverges, and one should evaluate the asymptotic behaviour of $\epsilon^{j-1}A_{j}^{\prime}$ to add to (\ref{a15}) and, therefore, characterize $\mathcal{L}[f](s)$ for $s\sim 2\hat J(k_{c})$ (or $\epsilon\sim 0$). Using $\cos x\leq 1-\frac{x^{2}}{\pi^{2}}$ for $x\in[0,\pi]\subset\mathbb{R}$,

\begin{eqnarray}
\nonumber \left|\epsilon^{j-1}A_{j}^{\prime}\right| & \leq & \left| \epsilon^{j-1}\int\limits_{B_{\delta}}\frac{d^{d}k}{\left(2\pi\right)^{d}}\frac{1}{\left[4Jd-4J\left(d-\sum_{i=1}^{d}\frac{k_{i}^{2}}{\pi^{2}}\right)\right]^{j}} + \mathcal{O}(\delta^{2}) \right| \\
\nonumber & \leq & \frac{\epsilon^{j-1}}{\left(4\pi^{-2}J\right)^{j}\left(2\pi\right)^{d}} \frac{2\pi^{\frac{d}{2}}}{\Gamma\left(\frac{d}{2}\right)} \left| \int\limits_{\delta}^{\pi\sqrt{d}} dkk^{d-1-2j} \right| + \mathcal{O}(\delta^{2}) \\
\nonumber & \leq & \left\{
\begin{array}{lcl}
\frac{2\pi^{j}}{\left(4\pi^{-2} J\right)^{j} \left(2\pi\right)^{2j}\Gamma(j)} \epsilon^{j-1} \left|\ln\delta\right| + \mathcal{O}(\epsilon^{j-1}) + \mathcal{O}(\delta^{2}) & , & d=2j \\
 & & \\
\frac{2\pi^{\frac{d}{2}} (2\pi)^{-d} \delta^{d-2} }{\left(4\pi^{-2}J\right)^{j} \Gamma\left(\frac{d}{2}\right)\left(2j-d\right)} \left(\frac{\epsilon}{\delta^{2}}\right)^{j-1} + \mathcal{O}(\epsilon^{j-1}) + \mathcal{O}(\delta^{2}) & , & d<2j
\end{array}
\right.. \\
\label{a17}
\end{eqnarray}

\indent

Therefore,

\begin{eqnarray}
\sum_{j=1}^{\infty} A_{j}^{\prime}\epsilon^{j-1} = \left\{
\begin{array}{lcl}
\mathcal{O}(\delta^{-(2-d)})                      & , & d<2 \\
 & & \\
\mathcal{O}(-\ln\delta^{2})                       & , & d=2 \\
 & & \\
A_{1} + \mathcal{O}(\delta^{d-2})                 & , & 2<d<4 \\
 & & \\
A_{1} + \mathcal{O}(\delta^{2})                   & , & d=4 \\
 & & \\
A_{1} - A_{2}\epsilon + \mathcal{O}(\delta^{d-4}) & , & d>4
\end{array}
\right..
\label{a18}
\end{eqnarray}

\indent

Combining equations (\ref{a15}) and (\ref{a18}),

\begin{eqnarray}
\mathcal{L}[f](s) \sim \left\{
\begin{array}{lcl}
\left(8\pi J\right)^{- \frac{d}{2}} \Gamma\left(1-\frac{d}{2}\right)\epsilon^{- \frac{2-d}{2}} & , & d<2 \\
 & & \\
\left(8\pi J\right)^{-1} \left(-\ln\epsilon\right) & , & d=2 \\
 & & \\
A_{1} - \left(8\pi J\right)^{- \frac{d}{2}} \left|\Gamma\left(1-\frac{d}{2}\right)\right|\epsilon^{\frac{d-2}{2}} & , & 2<d<4 \\
 & & \\
A_{1} - \left(8\pi J\right)^{-2} \left(-\epsilon\ln\epsilon\right) & , & d=4 \\
 & & \\
A_{1} - A_{2}\epsilon & , & d>4
\end{array}
\right..
\label{a19}
\end{eqnarray}

\subsection*{Asymptotic behaviour of $\psi$ - general comments}

In this Section the asymptotic behaviour of $\psi$,

\begin{eqnarray}
\psi(t) = \frac{1}{2\pi i}\int\limits_{\sigma -i\infty}^{\sigma +i\infty}ds e^{st} \mathcal{L}[\psi](s), \qquad \mathcal{L}[\psi](s) = \frac{\mathcal{L}[f](s)}{1-2T\mathcal{L}[f](s)},
\label{a20}
\end{eqnarray}

\noindent
where $\sigma$ is larger than the real part of any pole of the integrand, will be evaluated. First, one should note that the Laplace transform of $f$,

\begin{eqnarray}
\mathcal{L}[f](s) = \int\limits_{[-\pi, \pi]^{d}} \frac{d^{d}k}{\left(2\pi\right)^{d}} \frac{1}{s-2\hat J(k)},
\label{a21}
\end{eqnarray}

\noindent
has a cut on $[\inf_{k\in[-\pi,\pi]^{d}}\{2\hat J(k)\}, 2\hat J(k_{c})]$ in the complex $s$-plane. Furthermore, $\mathcal{L}[f]$ is a monotonically decreasing function of $s$, ranging from $0$ to $\beta_{c}/2$ ($\beta_{c}/2$ being infinite in the absence of phase transition).

\subsection*{Asymptotic behaviour of $\psi$ - supercritical case}

Let $P:=\{y\in\mathbb{C}: y\mbox{ is pole of }e^{st}\mathcal{L}[\psi](s)\}$ and let $\overline{p}:=\sup_{y\in P}\{\mbox{Re }y\}$. By the monotonicity of $\mathcal{L}[f](s)$, the denominator of $\mathcal{L}[\psi](s)$ (see (\ref{a20})) runs the interval $[1-\frac{T}{T_{c}}, 1]$ reaching each point only one time. Therefore, in the supercritical dynamics, $T>T_{c}$, equation (\ref{a20}) has a single pole, denoted henceforth by $\tau_{p}^{-1}$.

\begin{center}
\epsfig{file = 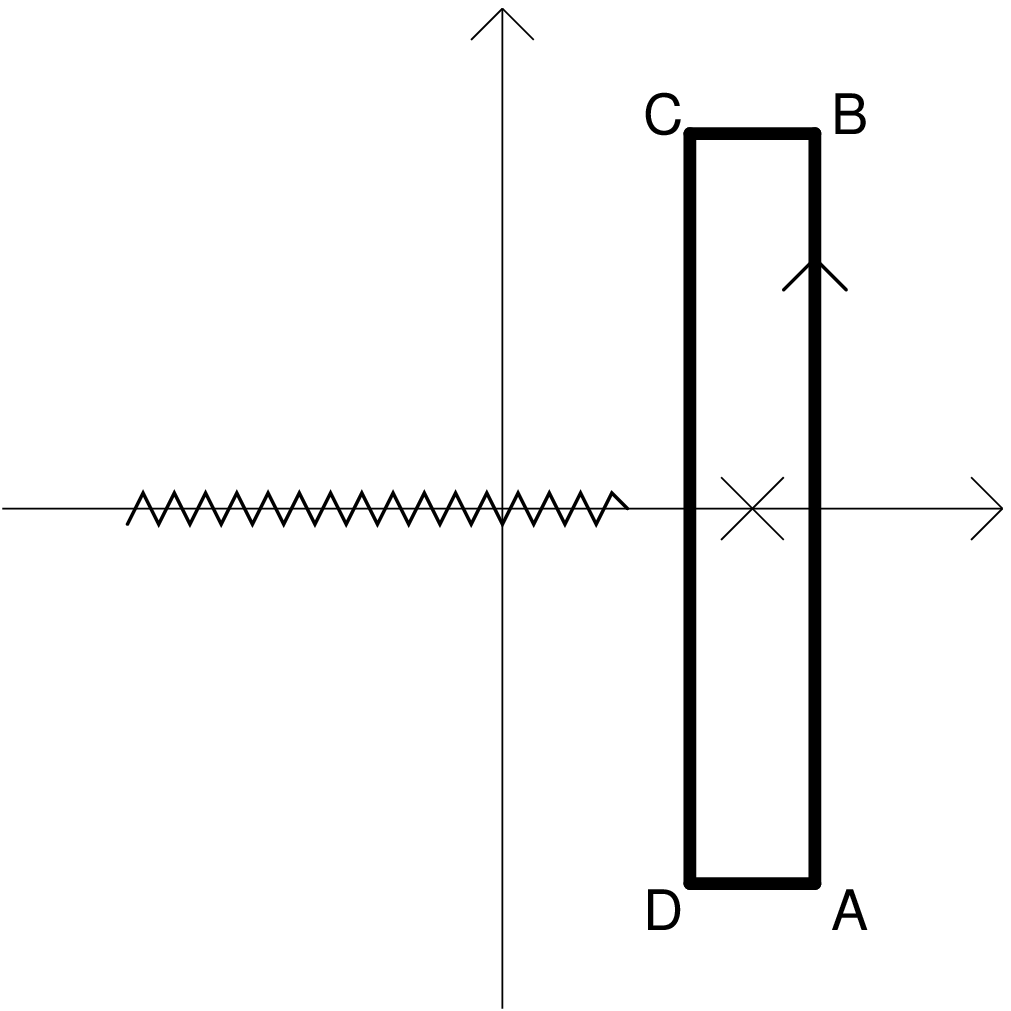, scale = 0.6}\\
\end{center}
\begin{center}
\textbf{Figure A1:} Contour for integration - supercritical case.
\end{center}

\indent

By the residue theorem, choosing the contour indicated in Figure A1, where $ABCD$ is a rectangle with vertices $c\pm iR$ and $\sigma\pm iR$ such that $2\hat J(k_{c})<c<\overline{p}<\sigma$,

\begin{eqnarray}
\nonumber 2\pi i \mbox{ Res } e^{st} \mathcal{L}[\psi](s) & = & \int\limits_{\sigma -iR}^{\sigma +iR}ds e^{st}\mathcal{L}[\psi](s) + e^{iRt}\int\limits_{\sigma}^{c}dy e^{yt} \mathcal{L}[\psi](y+iR) + \\
\nonumber & & + ie^{ct}\int\limits_{R}^{-R}dy e^{iyt}\mathcal{L}[\psi](c+iy) + \\
 & & + e^{-iRt}\int\limits_{c}^{\sigma}dy e^{yt}\mathcal{L}[\psi](y-iR).
\label{a22}
\end{eqnarray}

\indent

It is easy to see that $\lim_{R\rightarrow\infty}\left|\mathcal{L}[\psi](y\pm iR)\right|=0$. Moreover, the third term is $\mathcal{O}(e^{ct})$, which is negligible as compared with the first one (equal to $2\pi i\psi(t)$ in the limit $R\rightarrow\infty$), that is $\mathcal{O}(e^{\sigma t})$. Therefore,

\begin{eqnarray}
\psi(t) \sim \mbox{Res }e^{st}\mathcal{L}[\psi](s) = - \frac{1}{4T^{2}}\frac{1}{\partial_{s}\mathcal{L}[\psi](\tau_{p}^{-1})} e^{\frac{t}{\tau_{p}}}.
\label{a23}
\end{eqnarray}

\indent

As the temperature gets close to $T_{c}$ (from above), $\tau_{p}^{-1}$ became closer to $2\hat J(k_{c})$ (hitting this point at $T_{c}$), which is one of the edge of the cut in the complex $s$-plane. Hence, in the vicinity of $T_{c}$, one has

\begin{eqnarray}
\mbox{Res }e^{st}\mathcal{L}[\psi](s) \sim \lim_{s\sim 2\hat J(k_{c})^{+}} \left[s-2\hat J(k_{c})\right] e^{st}\mathcal{L}[\psi](s), \qquad T\sim T_{c}^{+},
\label{a24}
\end{eqnarray}

\noindent
and $\mathcal{L}[\psi]$ can be replaced by its asymptotic formula

\begin{eqnarray}
\mathcal{L}[\psi](s)\sim\left\{
\begin{array}{lcl}
\frac{F_{p}g_{p}}{\epsilon^{-\alpha_{p}}-2TF_{p}g_{p}} & , & d<d_{c} \\
 & & \\
\frac{F_{p}\left(-\ln\epsilon\right)}{1-2TF_{p}\left(-\ln\epsilon\right)} & , & d=d_{c} \\
 & & \\
\frac{A_{1}-F_{p}g_{p}\epsilon^{\alpha_{p}}}{1-2TA_{1}+2TF_{p}|g_{p}|\epsilon^{\alpha_{p}}} & , & d_{c}<d<\overline{d} \\
 & & \\
\frac{A_{1}-F_{p}\left(-\epsilon\ln\epsilon\right)}{1-2TA_{1}+2TF_{p}\left(-\epsilon\ln\epsilon\right)} & , & d=\overline{d} \\
 & & \\
\frac{A_{1}-A_{2}\epsilon}{1-2TA_{1}+2TA_{2}\epsilon} & , & d>\overline{d}
\end{array}
\right.,
\label{a25}
\end{eqnarray}

\noindent
which can be calculated from (\ref{Lf}) and (\ref{a20}). If these results are inserted in (\ref{a24}), one finds

\begin{eqnarray}
\tau_{p}^{-1}\sim\left\{
\begin{array}{lcl}
2\hat J(k_{c}) + \left(2TF_{p}g_{p}\right)^{- \frac{1}{\alpha_{p}}} & , & d<d_{c} \\
 & & \\
2\hat J(k_{c}) + \exp\left(- \frac{1}{2TF_{p}}\right) & , & d=d_{c} \\
 & & \\
2\hat J(k_{c}) + \left[-\frac{1}{2TF_{p}|g_{p}|}\left(1-\frac{T}{T_{c}}\right)\right]^{\frac{1}{\alpha_{p}}}& , & d_{c}<d<\overline{d} \\
 & & \\
2\hat J(k_{c}) + \epsilon^{\ast} & , & d=\overline{d} \\
 & & \\
2\hat J(k_{c}) + \left(-\frac{1}{2TA_{2}}\right)\left(1-\frac{T}{T_{c}}\right) & , & d>\overline{d} \\
 & &
\end{array}
\right.,
\label{a26}
\end{eqnarray}

\noindent
where $\epsilon^{\ast}$ is the least root of

\begin{eqnarray}
\epsilon\ln\epsilon=\frac{1}{2TF_{p}}\left(1-\frac{T}{T_{c}}\right).
\label{a27}
\end{eqnarray}

\indent

The characteristic relaxation time, $\tau_{eq}$, is related to $\tau_{p}$ by

\begin{eqnarray}
\tau_{eq}^{-1}=\tau_{p}^{-1}-2\hat J(k_{c}).
\label{a28}
\end{eqnarray}

\subsection*{Asymptotic behaviour of $\psi$ - critical case}

The simple pole of $\mathcal{L}[\psi](s)$, which is isolated in the supercritical case, touches $2\hat J(k_{c})$ (one of the edges of the cut) at $T=T_{c}$. Taking the integration contour as in figure A2, it is easy to show that

\begin{eqnarray}
\int\limits_{\sigma-iR}^{\sigma+iR} ds e^{st} \mathcal{L}[\psi](s) = \int\limits_{GFEDC} ds e^{st} \mathcal{L}[\psi](s),
\label{a29}
\end{eqnarray}

\noindent
since the contribution of the paths $BC$ and $GA$ vanishes in the limit $R\rightarrow\infty$.

\begin{center}
\epsfig{file = 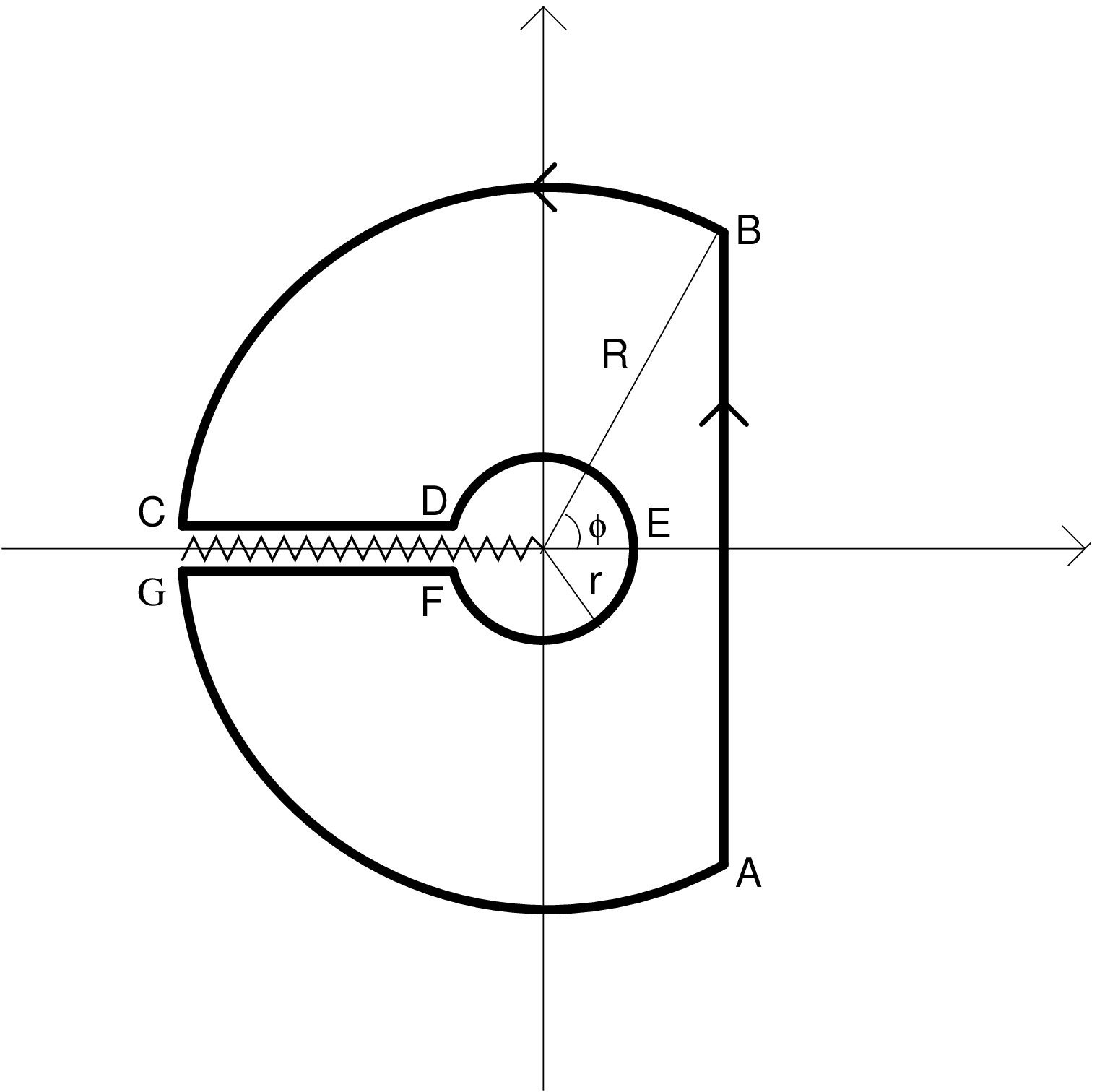, scale = 0.6}\\
\end{center}
\begin{center}
\textbf{Figure A2:} Contour for integration - critical case.
\end{center}

\indent

For sufficiently large time the integral in (\ref{a29}) is dominated by the contribution due to the path $FED$. Therefore, substituting $\mathcal{L}[f](s)$ by its asymptotic form (\ref{Lf}) is a suitable operation to evaluate the asymptotic form of $\psi$.

If $d_{c}<d<\overline{d}$, one has $\mathcal{L}[f](s)\sim A_{1} - F_{p}|g_{p}|\epsilon^{\alpha_{p}}$. Hence, $\mathcal{L}[\psi](s)\sim\frac{A_{1}^{2}}{F_{p}|g_{p}|}\epsilon^{-\alpha_{p}}$ and

\begin{eqnarray}
\psi(t) \sim \frac{A_{1}^{2}}{F_{p}|g_{p}|} \frac{1}{2\pi i} \int\limits_{GFEDC}ds e^{st} \left[s-2\hat J(k_{c})\right]^{-\alpha_{p}}.
\label{a30}
\end{eqnarray}

\indent

The integral in (\ref{a30}) is a Gamma function with Hankel's contour. Therefore,

\begin{eqnarray}
\psi(t) \sim \frac{A_{1}^{2}}{F_{p}|g_{p}|} \frac{t^{\alpha_{p}-1}e^{2\hat J(k_{c})t}}{\Gamma(\alpha)}.
\label{a31}
\end{eqnarray}

\indent

For $d=\overline{d}$, $\mathcal{L}[f](s) \sim A_{1} - F_{p} (- \epsilon \ln \epsilon)$, and, therefore, $\mathcal{L}[\psi](s) \sim - \frac{A_{1}^{2}}{ F_{p} \epsilon \ln \epsilon }$. The integration contour is again the one shown in figure A2. Since the contribution due to the arc $FED$ is zero (letting its radius to zero), one finds

\begin{eqnarray}
\nonumber \psi(t) & \sim & - \frac{A_{1}}{2T_{c}F_{p}} \frac{e^{2 \hat J(k_{c})t}}{2 \pi i} \left[ \int \limits_{\infty}^{0} \frac{d(re^{- i \pi}) e^{tre^{- i \pi}}}{ re^{- i \pi} \left( \ln r - i \pi \right) } + \int \limits_{0}^{\infty} \frac{d(re^{i \pi}) e^{tre^{i \pi}}}{ re^{i \pi} \left( \ln r + i \pi \right) } \right] \\
 & = & \frac{A_{1}}{2T_{c}F_{p}} e^{2 \hat J(k_{c})t} \int \limits_{0}^{\infty} \frac{dr e^{- rt}}{r \left( \ln^{2}r + \pi^{2} \right)}.
\label{a32}
\end{eqnarray}

\indent

Adopting the change of variables $r \rightarrow e^{\pi r}$, and integrating by parts, one has

\begin{eqnarray}
\nonumber \psi(t) & \sim & \frac{A_{1}}{2T_{c}F_{p}} \frac{ e^{2 \hat J(k_{c})t} }{\pi} \int \limits_{- \infty}^{\infty} \frac{ dr \exp \left( - t e^{\pi r} \right) }{r^{2} + 1} \\
\nonumber & = & \frac{A_{1}}{2T_{c}F_{p}} \frac{ e^{2 \hat J(k_{c})t} }{\pi} \bigg\{ \exp \left.\left( - t e^{\pi r} \right) \tan^{- 1}r \right|_{- \infty}^{\infty} + \\
\nonumber & & + \pi t \int \limits_{- \infty}^{\infty} dr \tan^{- 1}r \exp \left( \pi r - te^{\pi r} \right)  \bigg\} \\
\nonumber & = & \frac{A_{1}}{2T_{c}F_{p}} \frac{ e^{2 \hat J(k_{c})t} }{\pi} \left\{ \frac{\pi}{2} - \int \limits_{0}^{\infty} du e^{- u} \tan^{- 1} \left[ \frac{1}{\pi} \ln \left( \frac{t}{u} \right) \right] \right\} \\
\nonumber & = & \frac{A_{1}}{2T_{c}F_{p}} \frac{ e^{2 \hat J(k_{c})t} }{\pi} \left\{ \frac{\pi}{2} - \int \limits_{0}^{\infty} du e^{- u} \left[ \frac{\pi}{2} - \frac{\pi}{\ln \left( \frac{t}{u} \right) } + \mathcal{O}(\ln^{- 3}t) \right] \right\} \\
 & = & \frac{A_{1}}{2T_{c}F_{p}} \frac {e^{2 \hat J(k_{c})t} }{\ln t} \left[ \int \limits_{0}^{\infty} du \frac{ e^{- u} }{ 1 - \frac{\ln u}{\ln t} } + \mathcal{O}(\ln^{- 4}t) \right].
\label{a33}
\end{eqnarray}

\noindent
The application $e^{- u}\left(1 - \frac{\ln u}{\ln t}\right)^{-1}$, as a function of $u$, is defined on the interval $(0, \infty)$ almost everywhere, and the integral

\begin{eqnarray}
A(t) := \int \limits_{0}^{\infty} du \frac{e^{-u}}{1 - \frac{\ln u}{\ln t}}
\label{a34}
\end{eqnarray}

\noindent
is understood as the Cauchy principal value

\begin{eqnarray}
A(t) := \mbox{v.p.} \int \limits_{0}^{\infty} du \frac{e^{-u}}{1 - \frac{\ln u}{\ln t}}.
\label{a35}
\end{eqnarray}

\noindent
In the next Section it will be demonstrated that $\lim_{t\rightarrow\infty}A(t)=1$. Using this result, one finds

\begin{eqnarray}
\psi(t) \sim \frac{A_{1}}{2T_{c}F_{p}} \frac{ e^{2\hat J(k_{c})t} }{\ln t}
\label{a36}
\end{eqnarray}

\noindent
for $d=\overline{d}$.

If $d>\overline{d}$, $\mathcal{L}[f](s) \sim A_{1}-A_{2}\epsilon$ and $\mathcal{L}[\psi](s) \sim \frac{A_{1}^{2}}{A_{2}\left( s - 2 \hat J(k_{c}) \right)}$. Using again the contour of figure A2, this leads to

\begin{eqnarray}
\psi(t) \sim \frac{A_{1}^{2}}{A_{2}} \frac{1}{2 \pi i} \int \limits_{JEFGK} ds \frac{e^{st}}{s - 2 \hat J(k_{c})}\,,
\label{a37}
\end{eqnarray}

\noindent
with the contribution due to the path $GF$ and $DC$ being negligible as compared to the one due to $FED$. Letting its radius to zero, one sees that

\begin{eqnarray}
\psi(t) \sim \frac{A_{1}^{2}}{A_{2}} e^{2\hat J(k_{c})t}.
\label{a38}
\end{eqnarray}

\indent

The results are summarized by

\begin{eqnarray}
\psi(t) \sim \left\{
\begin{array}{lcl}
\displaystyle\frac{A_{1}^{2}}{F_{p}|g|\Gamma(\alpha_{p})} \frac{e^{2 \hat J(k_{c}) t}}{t^{1 - \alpha_{p}}} & & d_{c} < d < \overline{d} \\
 & & \\
\displaystyle\frac{A_{1}}{2 T_{c} F_{p}} \frac{e^{2 \hat J(k_{c}) t}}{\ln t} & & d = \overline{d} \\
 & & \\
\displaystyle\frac{A_{1}^{2}}{A_{2}} e^{2 \hat J(k_{c}) t} & & d > \overline{d}
\end{array}
\right..
\label{a39}
\end{eqnarray}

\subsection*{Proof of $\lim_{t\rightarrow\infty}A(t)=1$}

Considering $t \gg e$, the integral

\begin{eqnarray}
A(t) =\mbox{v.p.} \int \limits_{0}^{\infty} du \frac{e^{- u}}{1 - \frac{\ln u}{\ln t}} = A^{(1)}(t) + A^{(2)}(t) + A^{(3)}(t)
\label{a40}
\end{eqnarray}

\noindent
is divided in three parts such that

\begin{eqnarray}
A^{(1)}(t) := \int \limits_{0}^{\frac{1}{\ln t}} du \frac{e^{- u}}{1 - \frac{\ln u}{\ln t}} = \ln t \int \limits_{0}^{\frac{1}{\ln t}} du \frac{e^{- u}}{\ln t + |\ln u|}.
\label{a41}
\end{eqnarray}

\indent

Therefore,

\begin{eqnarray}
\nonumber |A^{(1)}(t)| & \leq & \frac{\ln t}{\ln t + \ln\ln t} \int \limits_{0}^{\frac{1}{\ln t}} du e^{- u} = \frac{\ln t}{\ln t + \ln\ln t} \left( 1 - e^{- \frac{1}{\ln t}} \right) \\
 & = & \frac{\ln t}{\ln t + \ln\ln t} \left[ \frac{1}{\ln t} + \mathcal{O}(\ln^{- 2}t) \right],
\label{a42}
\end{eqnarray}

\noindent
and $\lim_{t \rightarrow \infty} A^{(1)}(t) = 0$.

The second term, $A^{(2)}(t)$, is responsible for the non zero contribution of the asymptotic behaviour of $A(t)$,

\begin{eqnarray}
A^{(2)}(t) := \int \limits_{\frac{1}{\ln t}}^{\ln t} du \frac{e^{- u}}{1 - \frac{\ln u}{\ln t}} = \int \limits_{\frac{1}{\ln t}}^{\ln t} du e^{- u} \left[ 1 + o(1) \right] = e^{- \frac{1}{\ln t}} - e^{- \ln t} + o(1)\,;
\label{a43}
\end{eqnarray}

\noindent
hence, $\lim_{t \rightarrow \infty}A^{(2)}(t)=1$.

It remains to show that the third term, $A^{(3)}(t)$, is zero for $t\rightarrow\infty$. Separating in three parts,

\begin{eqnarray}
A^{(3)}(t) := \mbox{v.p.} \int \limits_{\ln t}^{\infty} du \frac{e^{- u}}{1 - \frac{\ln u}{\ln t}} = A^{(3a)}(t) + A^{(3b)}(t) + A^{(3c)}(t),
\label{a44}
\end{eqnarray}

\noindent
it will be shown that each one of them vanishes in the limit $t\rightarrow\infty$.

The first term,

\begin{eqnarray}
A^{(3a)}(t) := \int \limits_{\ln t}^{t \left( 1 - \frac{1}{\ln t} \right)} du \frac{e^{- u}}{1 - \frac{\ln u}{\ln t}} = - t \ln t \int \limits_{\frac{\ln t}{t}}^{1 - \frac{1}{\ln t}} du \frac{e^{- tu}}{\ln u},
\label{a45}
\end{eqnarray}

\noindent
admits the bound

\begin{eqnarray}
\nonumber |A^{(3a)}(t)| & \leq & t \ln t \int \limits_{\frac{\ln t}{t}}^{1 - \frac{1}{\ln t}} du \frac{e^{- tu}}{|\ln u|} \leq \frac{t \ln t}{\left| \ln \left( 1 - \frac{1}{\ln t} \right) \right|} \int \limits_{\frac{\ln t}{t}}^{1 - \frac{1}{\ln t}} du e^{- tu} \\
 & = & \frac{t \ln t}{\left| \frac{1}{\ln t} + \mathcal{O} \left( \frac{1}{\ln^{2} t} \right) \right|} \frac{1}{t} \left[ \frac{1}{t} - e^{- t \left( 1 - \frac{1}{\ln t} \right)} \right],
\label{a46}
\end{eqnarray}

\noindent
from which $\lim_{t \rightarrow \infty}A^{(3a)}(t) = 0$.

The second term,

\begin{eqnarray}
A^{(3b)}(t) := \mbox{v.p.} \int \limits_{t \left( 1 - \frac{1}{\ln t} \right)}^{t \left( 1 + \frac{1}{\ln t} \right)} du \frac{e^{- u}}{1 - \frac{\ln u}{\ln t}} = - t \ln t \mbox{v.p.} \int \limits_{1 - \frac{1}{\ln t}}^{1 + \frac{1}{\ln t}} du \frac{e^{- tu}}{\ln u}\,,
\label{a47}
\end{eqnarray}

\noindent
can be bounded by

\begin{eqnarray}
\nonumber |A^{(3b)}(t)| & \leq & t \ln t \; \mbox{v.p.} \int \limits_{1 - \frac{1}{\ln t}}^{1 + \frac{1}{\ln t}} du \frac{e^{- tu}}{|\ln u|} \\
\nonumber & \leq & t e^{- t \left( 1 - \frac{1}{\ln t} \right)} \ln t \; \mbox{v.p.} \int \limits_{1 - \frac{1}{\ln t}}^{1 + \frac{1}{\ln t}} du \frac{1}{|\ln u|} \\
\nonumber & = & t e^{- t \left( 1 - \frac{1}{\ln t} \right)} \ln t \; \mbox{v.p.} \int \limits_{- \frac{1}{\ln t}}^{\frac{1}{\ln t}} \frac{du}{|\ln \left( 1 + u \right)|} \\
\nonumber & = & t e^{- t \left( 1 - \frac{1}{\ln t} \right)} \ln t \; \mbox{v.p.} \int \limits_{- \frac{1}{\ln t}}^{\frac{1}{\ln t}} \frac{du}{|u|} \left[ 1 + \frac{|u|}{2} + \mathcal{O}(u^{2}) \right] \\ 
 & = & t e^{- t \left( 1 - \frac{1}{\ln t} \right)} \ln t \left[ \frac{1}{\ln t} + o (\ln^{- 1} t) \right]\,,
\label{a48}
\end{eqnarray}

\noindent
and $\lim_{t \rightarrow \infty} A^{(3b)}(t) = 0$.

Finally, the third term,

\begin{eqnarray}
A^{(3c)}(t) := \int \limits_{t \left( 1 + \frac{1}{\ln t} \right)}^{\infty} du \frac{e^{- u}}{1 - \frac{\ln u}{\ln t}} = - t \ln t \int \limits_{1 + \frac{1}{\ln t}}^{\infty} du \frac{e^{- tu}}{\ln u}\,,
\label{a49}
\end{eqnarray}

\noindent
also goes to zero in the limit $t \rightarrow \infty$,

\begin{eqnarray}
|A^{(3c)}(t)| \leq \frac{t \ln t}{\ln \left( 1 + \frac{1}{\ln t} \right)} \int \limits_{1 + \frac{1}{\ln t}}^{\infty} du e^{- tu} = \frac{t \ln t}{\frac{1}{\ln t} + \mathcal{O}(\ln^{- 2}t)} \frac{e^{- t \left( 1 + \frac{1}{\ln t} \right)}}{t},
\label{a50}
\end{eqnarray}

\noindent
which implies $\lim_{t \rightarrow \infty} A^{(3c)}(t) = 0$, and completes the proof.

\subsection*{Asymptotic behaviour of $\psi$ - subcritical case}

In the subcritical case, where $\mathcal{L}[f](s) \sim A_{1} - h \left( \left[ s - 2 \hat J(k_{c} \right] ) \right)$ for $s\sim 2\hat J(k_{c})$, $h$ being a continuous function for $s \geq 2 \hat J(k_{c})$, and $h(0) = 0$. The form of $h$ is given in (\ref{Lf}). It is easily seen that

\begin{eqnarray}
\nonumber \mathcal{L}[\psi](s) = \frac{1}{M_{eq}^{4}} \mathcal{L}[f](s) - \frac{TA_{1}}{M_{eq}^{4} T_{c}} + \mathcal{O} \left( \left[h(s - 2 \hat J(k_{c}) \right]^{2} \right), \quad s \sim 2 \hat J(k_{c}),\\
\label{a51}
\end{eqnarray}

\noindent
where

\begin{eqnarray}
M_{eq}^{2} := 1 - \frac{T}{T_{c}}.
\label{a52}
\end{eqnarray}

\noindent
In analogy to the manipulations in the critical case, and using the contour given in figure A2, one finds, for large times, that

\begin{eqnarray}
\psi(t) = \frac{1}{M_{eq}^{4}} \int \limits_{GFEDC} ds e^{st} \mathcal{L}[f](s) + \mathcal{O} \left( \left[h(s - 2 \hat J(k_{c}) \right]^{2} \right).
\label{a53}
\end{eqnarray}

\indent

The integral in equation (\ref{a53}) can be written as

\begin{eqnarray}
\psi(t) \sim \frac{1}{M_{eq}^{4}} \mathcal{L}^{- 1} \mathcal{L} [f](t) = \frac{f(t)}{M_{eq}^{4}}.
\label{a54}
\end{eqnarray}

\subsection*{Three auxiliary equations}

The following three equations are useful in the calculations of the asymptotic behaviour of the two-time functions (autocorrelation and response function):

\begin{eqnarray}
\mbox{(i)} \quad \frac{1}{2T_{c}} = \int\limits_{0}^{\infty} dt e^{-2\hat J(k_{c})t} f(t).
\label{a55}
\end{eqnarray}

\noindent
This equation is the definition of critical temperature, and it is easily obtained by the definition of $f$.

\begin{eqnarray}
\mbox{(ii)} \quad \frac{1}{2T} = \int\limits_{0}^{\infty} dt e^{- \frac{t}{\tau_{p}}} f(t) \quad (T>T_{c}).
\label{a56}
\end{eqnarray}

\noindent
From equation (\ref{a20}), and $\tau_{p}^{-1}$ being the simple pole of $\mathcal{L}[\psi](s)$, one has

\begin{eqnarray}
\mathcal{L}[f](\tau_{p}^{-1}) = \frac{1}{2T}.
\label{a57}
\end{eqnarray}

\noindent
Therefore,

\begin{eqnarray}
\int \limits_{0}^{\infty} dt e^{- \frac{t}{\tau_{p}}} f(t) = \lim_{s \rightarrow \tau_{p}^{- 1}} \int \limits_{0}^{\infty} dt e^{- st} f(t) = \lim_{s \rightarrow \tau_{p}^{- 1}} \mathcal{L}[f](s) = \frac{1}{2T},
\label{a58}
\end{eqnarray}

\noindent
which is the desired result.

\begin{eqnarray}
\mbox{(iii)} \quad \frac{1}{2T_{c}M_{eq}^{2}} = \int\limits_{0}^{\infty} dt e^{-2\hat J(k_{c})t} \psi(t) \quad (T<T_{c}).
\label{a59}
\end{eqnarray}

\noindent
For $T<T_{c}$, one sees that $\mathcal{L}[f](s) \sim A_{1} + h(s - 2 \hat J(k_{c}))$ for $s \sim 2 \hat J(k_{c})^{+}$, where $h$ is a continuous application of $s - 2 \hat J (k_{c})$, and therefore, $h(0) = 0$. Hence, by this expansion and from equation (\ref{a20}),

\begin{eqnarray}
\nonumber \lim_{s \rightarrow 2 \hat J(k_{c})^{+}} \int \limits_{0}^{\infty} dt e^{- st} \psi(t) & = & \lim_{s \rightarrow 2 \hat J(k_{c})^{+}} \mathcal{L}[\psi](s) \\
 & = & \lim_{s \rightarrow 2 \hat J(k_{c})^{+}} \frac{\mathcal{L}[f](s)}{1 - 2T \mathcal{L}[f](s)} = \frac{A_{1}}{1 - 2TA_{1}}.
\label{a60}
\end{eqnarray}

\noindent
The desired result follows from $A_{1} = \frac{1}{2T_{c}}$.

\subsection*{Autocorrelation, response function, and fluctuation - dissipation ratio}

The autocorrelation (\ref{Cpsi}),

\begin{eqnarray}
\nonumber C(t, t^{\prime}) & = & \frac{1}{ \sqrt{ \psi(t) \psi(t^{\prime}) } } \left[ f \left( \frac{t + t^{\prime}}{2} \right) + 2T \int \limits_{0}^{t^{\prime}} dy f \left( \frac{t + t^{\prime}}{2} - y\right) \psi(y) \right] \\
\nonumber & = & \frac{1}{ \sqrt{ \psi( t^{\prime} + \tau ) \psi(t^{\prime}) } } \left[ f \left( t^{\prime} + \frac{\tau}{2} \right) + 2T \int \limits_{0}^{t^{\prime}} dy f \left( t^{\prime} - y + \frac{\tau}{2} \right) \psi(y) \right], \\
\label{a61}
\end{eqnarray}







\noindent
the response function,

\begin{eqnarray}
R(t, t^{\prime}) = f \left( \frac{\tau}{2} \right) \sqrt{\frac{ \psi(t^{\prime} )}{ \psi(t) }},
\label{a64}
\end{eqnarray}

\noindent
and the fluctuation-dissipation ratio,

\begin{eqnarray}
X(t, t^{\prime}) = \frac{T R(t, t^{\prime})}{\partial_{t^{\prime}} C(t, t^{\prime})},
\label{a65}
\end{eqnarray}

\noindent
display different behaviours, that depend on the temperature and the chosen time scale. The calculations of these functions will be divided in three parts, each of them corresponding to different choices of temperature. In each part, distinct time scales, leading to the dynamical behaviour of the two-time function, will be considered. The notation

\begin{eqnarray}
x := \frac{t}{t^{\prime}}
\label{a66}
\end{eqnarray}

\noindent
will be used.

Let $\psi_{a}$ be the asymptotic form of $\psi$. In other words,

\begin{eqnarray}
\psi(t) \sim \psi_{a}(t) = \left\{
\begin{array}{lcl}
D_{>} e^{t/\tau_{p}} & , & T > T_{c} \\
 & & \\
D_{1=} \frac{e^{2\hat J(k_{c})t}}{t^{2-\gamma_{p}}} & , & T = T_{c} \quad \mbox{and} \quad d_{c}<d<\overline{d} \\
 & & \\
D_{2=} \frac{e^{2\hat J(k_{c})t}}{\ln t} & , & T = T_{c} \quad \mbox{and} \quad d=\overline{d} \\
 & & \\
D_{3=} e^{2\hat J(k_{c})t} & , & T = T_{c} \quad \mbox{and} \quad d>\overline{d} \\
 & & \\
\frac{f(t)}{M_{eq}^{4}} & , & T < T_{c}
\end{array}
\right.\,.
\label{a67}
\end{eqnarray}

Choosing an $\epsilon>0$ such that $1\ll \epsilon t^{\prime}\ll t^{\prime}$, one can write (\ref{a61}) as

\begin{eqnarray}
\nonumber C(t, t^{\prime}) & \sim & \frac{1}{\sqrt{\psi_{a}(t)\psi_{a}(t^{\prime})}} \left[ \frac{K_{p} e^{\hat J(k_{c})(2t^{\prime}+\tau)}}{\left(t^{\prime}+\frac{\tau}{2}\right)^{\gamma_{p}}} + 2T \hspace{-1.16mm} \int\limits_{0}^{\epsilon t^{\prime}} \hspace{-1.17mm} dy \frac{K_{p} e^{\hat J(k_{c})(2t^{\prime}+\tau-2y) }}{\left(t^{\prime}+\frac{\tau}{2}-y\right)^{\gamma_{p}}} \psi(y) + \right. \\
\nonumber & & \left. + 2T \int\limits_{\epsilon t^{\prime}}^{t^{\prime}} dy f\left(t^{\prime}+\frac{\tau}{2}-y\right) \psi_{a}(y) \right] \\
\nonumber & = & \frac{1}{\sqrt{\psi_{a}(t)\psi_{a}(t^{\prime})}} \left[ \frac{K_{p} e^{\hat J(k_{c})(2t^{\prime}+\tau)}}{\left(t^{\prime}+\frac{\tau}{2}\right)^{\gamma_{p}}} + \frac{2TK_{p}e^{\hat J(k_{c})(2t^{\prime}+\tau)}}{\left(t^{\prime}+\frac{\tau}{2}\right)^{\gamma_{p}}} \times \right. \\
\nonumber & & \left. \times \int\limits_{0}^{\epsilon t^{\prime}} dy e^{- \hat 2 \hat J(k_{c})y}\psi(y) + 2T \int\limits_{\frac{\tau}{2}}^{\left(1-\epsilon\right) t^{\prime}+\frac{\tau}{2}} dy f(y) \psi_{a} \left( t^{\prime} + \frac{\tau}{2} - y\right) \right]\,. \\
\label{a68}
\end{eqnarray}

\noindent
Since the function $w(y)=\psi(y)e^{-2\hat J(k_{c})y}$ is positive and non - increasing ($dw(y)/dy \leq 0$) on the real line, equation (\ref{a68}), can be written as

\begin{eqnarray}
\nonumber C(t, t^{\prime}) & \sim & \frac{1}{\sqrt{\psi_{a}(t) \psi_{a}(t^{\prime})}} \left[ \mathcal{O} \left( \frac{\epsilon t^{\prime} e^{\hat J(k_{c})(2t^{\prime}+\tau)} }{ (t^{\prime}+\tau/2)^{\gamma_{p}} } \right) + \right. \\
 & & \left. + 2T \int\limits_{\frac{\tau}{2}}^{\left(1-\epsilon\right)t^{\prime}+\frac{\tau}{2}} dy f(y) \psi_{a} \left( t^{\prime} + \frac{\tau}{2} - y\right) \right]\,.
\label{a69}
\end{eqnarray}

\subsection*{Supercritical dynamics}

By equations (\ref{a67}) and (\ref{a69}), one finds

\begin{eqnarray}
\nonumber C(t, t^{\prime}) & \sim & \mathcal{O} \left( \frac{\epsilon t^{\prime} e^{(\hat J(k_{c}) - \frac{1}{2\tau_{p}})(2t^{\prime}+\tau)} }{ \left(t^{\prime}+\frac{\tau}{2}\right)^{\gamma_{p}} } \right) + 2T \int\limits_{\frac{\tau}{2}}^{(1-\epsilon)t^{\prime}+\tau/2} dy f(y) e^{-y/\tau_{p}} \\
\nonumber & = & \mathcal{O} \left( \frac{\epsilon t^{\prime} e^{(\hat J(k_{c}) - \frac{1}{2\tau_{p}})(2t^{\prime}+\tau)} }{ \left(t^{\prime}+\frac{\tau}{2}\right)^{\gamma_{p}} } \right) + 2T \int\limits_{\frac{\tau}{2}}^{\infty} dy f(y) e^{-y/\tau_{p}} - \\
 & & - 2T \int\limits_{(1-\epsilon)t^{\prime}+\tau/2}^{\infty} dy f(y) e^{-y/\tau_{p}}\,.
\label{a70}
\end{eqnarray}

\noindent
In the asymptotic limit $t^{\prime}\sim\infty$, the first and third terms are negligible as compared with the second one, which is $\mathcal{O}(1)$ (see (\ref{a56})). Therefore, one has

\begin{eqnarray}
C(t, t^\prime{}) \sim T \int\limits_{\tau}^{\infty} dy f\left(\frac{y}{2}\right) e^{- \frac{y}{2\tau_{p}}}\,.
\label{a71}
\end{eqnarray}

By equations (\ref{a23}) and (\ref{a64}), the response function is

\begin{eqnarray}
R(t, t^{\prime}) \sim f \left( \frac{\tau}{2} \right) e^{- \frac{ \tau}{ 2 \tau_{p} }}.
\label{a72}
\end{eqnarray}

\indent

Using equations (\ref{a71}) and (\ref{a72}), one checks the fluctuation-dissipation theorem,

\begin{eqnarray}
X(t, t^{\prime}) \sim T f \left( \frac{\tau}{2} \right) e^{- \frac{ \tau }{ 2 \tau_{p} }} \left[ - \frac{\partial}{\partial \tau} T \int \limits_{\tau}^{\infty} dy f \left( \frac{y}{2} \right) e^{- \frac{y}{ 2\tau_{p}} } \right]^{- 1} = 1.
\label{a73}
\end{eqnarray}

\subsection*{Critical dynamics}

In the stationary regime ($1\sim \tau \ll t^{\prime }$), one may rewrite (\ref{a69}) as 

\begin{eqnarray}
\nonumber C(t, t^{\prime}) & \sim & \frac{1}{\sqrt{\psi_{a}(t)\psi_{a}(t^{\prime})}} \left[ \mathcal{O} \left( \frac{\epsilon t^{\prime} e^{\hat J(k_{c})(2t^{\prime}+\tau)} }{ \left(t^{\prime}+\frac{\tau}{2}\right)^{\gamma_{p}} } \right) + 2T_{c} \int\limits_{\frac{\tau}{2}}^{\epsilon t^{\prime}+\tau/2} dy f(y) \times \right. \\
\nonumber & & \left. \times \psi_{a}\left( t^{\prime} + \frac{\tau}{2} - y \right) + 2T_{c} \int\limits_{\epsilon t^{\prime}+\tau/2}^{(1-\epsilon) t^{\prime}+\tau/2} dy f(y) \psi_{a}\left( t^{\prime} + \frac{\tau}{2} - y \right) \right]\,, \\
\label{a74}
\end{eqnarray}

\noindent
which may be convenient for calculating the analytic asymptotic form of the autocorrelation in this regime.

\subsubsection*{Critical dynamics ($d_{c}<d<\overline{d}$)}

\medskip

By (\ref{a67}) and (\ref{a74}), in the stationary regime, one has

\begin{eqnarray}
\nonumber C(t, t^{\prime}) & \sim & \left[\left( t^{\prime}+\tau \right) t^{\prime}\right]^{\frac{2-\gamma_{p}}{2}} \left[ \mathcal{O}\left( \frac{\epsilon t^{\prime}}{ (t^{\prime}+\tau/2)^{\gamma_{p}} }\right) + \frac{2T_{c}}{\left( t^{\prime}+ \frac{\tau}{2}\right)^{2-\gamma_{p}}} \int\limits_{\frac{\tau}{2}}^{\epsilon t^{\prime}+\tau/2} dy \times \right. \\ 
\nonumber & & \left. \times f(y) e^{-2\hat J(k_{c})y} + 2T_{c} K_{p} \int\limits_{\epsilon t^{\prime}+\tau/2}^{(1-\epsilon)t^{\prime}+\tau/2} \frac{dy}{y^{\gamma_{p}} \left(t^{\prime} + \frac{\tau}{2} - y\right)^{2-\gamma_{p}}} \right]
\label{a75}
\end{eqnarray}

\noindent
Performing the change of variable $y\rightarrow 1/y$ in the last term, one finds

\begin{eqnarray}
\nonumber C(t, t^{\prime}) & \sim & \left[\left(t^{\prime}+\tau\right) t^{\prime} \right]^{\frac{2-\gamma_{p}}{2}} \left[ \mathcal{O} \left( \frac{\epsilon t^{\prime} }{ \left(t^{\prime}+\frac{\tau}{2}\right)^{\gamma_{p}} }\right) + \frac{1}{\left(t^{\prime}+\frac{\tau}{2}\right)^{2-\gamma_{p}}} C_{eq, c}(\tau)\right]\,, \\
\label{a76}
\end{eqnarray}

\noindent
where $C_{eq, c}$ is defined as

\begin{eqnarray}
C_{eq, c}(\tau) = T_{c} \int\limits_{\tau}^{\infty} dy f\left(\frac{y}{2}\right)e^{-\hat J(k_{c})y}
\label{ceq}
\end{eqnarray}

Restricting the range of $\epsilon>0$ to be such that $1\ll\epsilon t^{\prime}\ll (t^{\prime})^{\min\{1, 2\gamma_{p}-2\}}$ leads to

\begin{eqnarray}
C(t, t^{\prime}) \sim C_{eq, c}(\tau)\,.
\label{a77}
\end{eqnarray}

Since the response function is

\begin{eqnarray}
\nonumber R(t, t^{\prime}) & \sim & f \left( \frac{\tau}{2} \right) \sqrt{ \frac{e^{2 \hat J(k_{c}) t^{\prime}}/t^{\prime^{1 - \alpha_{p}}}}{ e^{2 \hat J(k_{c}) t}/t^{1 - \alpha_{p}} } } \\
\nonumber & = & f \left( \frac{\tau}{2} \right) \frac{ e^{- \hat J(k_{c}) \tau} }{ \left( 1 + \frac{\tau}{t^{\prime}} \right)^{1 - \alpha_{p}}} \\
 & \sim & f \left( \frac{\tau}{2} \right) e^{- \hat J(k_{c}) \tau},
\label{a78}
\end{eqnarray}

\noindent
it is also a function of $\tau$ only, and the fluctuation - dissipation theorem, $X(t, t^{\prime})\sim 1$, is asymptotically obeyed.

In the aging regime, from (\ref{a69}), one has

\begin{eqnarray}
\nonumber C(t, t^{\prime}) & \sim & \left[\left(t^{\prime}+\tau\right) t^{\prime}\right]^{\frac{2-\gamma_{p}}{2}} \left[ \mathcal{O}\left( \frac{\epsilon t^{\prime}}{ \left(t^{\prime}+\frac{\tau}{2}\right)^{\gamma_{p}} } \right) + 2T_{c} K_{p} \times \right. \\
\nonumber & & \left. \times \int\limits_{\frac{\tau}{2}}^{(1-\epsilon)t^{\prime}+\tau/2} \frac{dy}{ y^{\gamma_{p}} \left( t^{\prime}+\frac{\tau}{2}-y\right)^{2-\gamma_{p}}} \right] \\
\nonumber & \sim & \left[\left(t^{\prime}+\tau\right) t^{\prime}\right]^{\frac{2-\gamma_{p}}{2}} \left[ \mathcal{O}\left( \frac{\epsilon t^{\prime} }{ \left(t^{\prime}+\frac{\tau}{2}\right)^{\gamma_{p}} } \right) + \right. \\
 & & \left. + \frac{2T_{c} K_{p}}{\left(\gamma_{p}-1\right)\left(t^{\prime}+\frac{\tau}{2}\right) } \left(\frac{2t^{\prime}}{\tau}\right)^{\gamma_{p}-1} \right]\,.
\label{a79}
\end{eqnarray}

\noindent
Restricting $\epsilon>0$ to be such that $1\ll\epsilon (t^{\prime}+\tau/2)\ll (t^{\prime}/\tau)^{\gamma_{p}-1}(t^{\prime}+\tau/2)^{\gamma_{p}}$, one finds

\begin{eqnarray}
C(t, t^{\prime}) \sim \frac{2 K_{p} T_{c} 2^{\gamma_{p}}}{\gamma_{p} - 1} t^{\prime^{1 - \gamma_{p}}} \frac{ x^{1 - \frac{\gamma_{p}}{2}} \left( x - 1 \right)^{1 - \gamma_{p}} }{ x + 1 }.
\label{a80}
\end{eqnarray}

From equations (\ref{fasymp}) and (\ref{a64}), one calculates the response function,

\begin{eqnarray}
\nonumber R(t, t^{\prime}) & \sim & f \left( \frac{\tau}{2} \right) \sqrt{ \frac{e^{2 \hat J(k_{c}) t^{\prime}}/t^{\prime^{1 - \alpha_{p}}}}{e^{2 \hat J(k_{c}) t}/t^{1 - \alpha_{p}}} } \\
\nonumber & = & f \left( \frac{\tau}{2} \right) e^{- \hat J(k_{c}) \tau} x^{\frac{1 - \alpha_{p}}{2}} \\
 & \sim & 2^{\gamma_{p}} K_{p} t^{\prime^{- \gamma_{p}}} \left( x - 1 \right)^{- \gamma_{p}} x^{\frac{1 - \alpha_{p}}{2}},
\label{a81}
\end{eqnarray}

\noindent
and the fluctuation-dissipation ratio,

\begin{eqnarray}
X(t, t^{\prime}) \sim \frac{ \left( \gamma_{p} - 1 \right) \left( x + 1 \right)^{2} }{ \left( \gamma_{p} x + \gamma_{p} - 2 \right) \left( x + 1 \right) - 2 \left( x - 1 \right) },
\label{a82}
\end{eqnarray}

\noindent
which comes from equations (\ref{a80}) and (\ref{a81}).


\subsubsection*{Critical dynamics ($d=\overline{d}$)}

\medskip

In the stationary regime, $1 \sim \tau \ll t^{\prime}$, for $d=\overline{d}$, one should proceed analogously as was done in the stationary regime of case $d_{c}<d<\overline{d}$. Therefore, by choosing $\epsilon>0$ such that $1\ll \epsilon t^{\prime}\ll (t^{\prime}+\tau/2)^{\gamma_{p}}/\ln (t^{\prime}+\tau/2)$, one finds, by equations (\ref{a67}) and (\ref{a74}), that

\begin{eqnarray}
\nonumber C(t, t^{\prime}) & \sim & \sqrt{\ln(t^{\prime}+\tau)\ln t^{\prime}} \left[ \mathcal{O} \left( \frac{\epsilon t^{\prime}}{ \left(t^{\prime}+\frac{\tau}{2}\right)^{\gamma_{p}}} \right) + \frac{2T_{c}}{\ln\left( t^{\prime}+\frac{\tau}{2} \right)} \times \right. \\
\nonumber & & \left. \times \int\limits_{\frac{\tau}{2}}^{\epsilon t^{\prime}+\tau/2} dy f(y) e^{-2\hat J(k_{c})y} + 2T_{c}K_{p} \int\limits_{\epsilon t^{\prime}+\tau/2}^{(1-\epsilon)t^{\prime}+\tau/2} \frac{dy}{y^{\gamma_{p}}\ln\left(t^{\prime}+\frac{\tau}{2}-y\right)} \right] \\
\nonumber & = & \sqrt{\ln(t^{\prime}+\tau)\ln t^{\prime}} \left[ \mathcal{O} \left( \frac{\epsilon t^{\prime}}{ \left(t^{\prime}+\frac{\tau}{2}\right)^{\gamma_{p}}} \right) + \frac{2T_{c}}{\ln\left( t^{\prime}+\frac{\tau}{2} \right)} \times \right. \\
\nonumber & & \times \left( \int\limits_{\frac{\tau}{2}}^{\infty} dy f(y) e^{-2\hat J(k_{c})y} - \int\limits_{\epsilon t^{\prime}+\tau/2}^{\infty} \frac{K_{p} dy}{y^{\gamma_{p}}} \right) + \\
\nonumber & & \left. + \mathcal{O} \left( \frac{1}{\epsilon t^{\prime}} \int\limits_{\epsilon t^{\prime}+\tau/2}^{(1-\epsilon)+\tau/2} \frac{dy}{y^{\gamma_{p}}}\right) \right] \\
 & \sim & C_{eq, c}(\tau)\,.
\label{a83}
\end{eqnarray}

\noindent
Furthermore, by equation (\ref{a64}), it can be shown that

\begin{eqnarray}
\nonumber R(t, t^{\prime}) & \sim & f \left( \frac{\tau}{2} \right) \sqrt{ \frac{e^{2 \hat J(k_{c}) t^{\prime}}/\ln t^{\prime}}{e^{2 \hat J(k_{c}) t}/\ln \left( t^{\prime} + \tau \right)} } \\
\nonumber & = & f \left( \frac{\tau}{2} \right) \frac{ e^{- \hat J(k_{c}) \tau} }{ \sqrt{ 1 + \frac{\ln \left( 1 + \tau/t^{\prime} \right)}{\ln t^{\prime}}} } \\
 & \sim & f \left( \frac{\tau}{2} \right) e^{- \hat J(k_{c}) \tau},
\label{a84}
\end{eqnarray}

\noindent
ensuring that $X(t, t^{\prime}) \sim 1$ in the stationary time scale.

In the aging regime, $1 \ll \tau \sim t^{\prime}$, from (\ref{a67}) and (\ref{a69}), one finds

\begin{eqnarray}
\nonumber C(t, t^{\prime}) & \sim & \sqrt{\ln t \ln t^{\prime}} \left[ \mathcal{O} \left(\frac{\epsilon t^{\prime}}{\left(t^{\prime}+\frac{\tau}{2}\right)^{\gamma_{p}}}\right) + 2T_{c}K_{p} \hspace{-2.49mm}\int\limits_{\frac{\tau}{2}}^{(1-\epsilon)t^{\prime}+\tau/2} \hspace{-2.50mm} \frac{dy}{ y^{\gamma_{p}}\ln\left(t^{\prime}+\frac{\tau}{2}-y\right) }\right] \\
\nonumber & = & \sqrt{\ln t \ln t^{\prime}} \left[ \mathcal{O} \left(\frac{\epsilon t^{\prime}}{\left(t^{\prime}+\frac{\tau}{2}\right)^{\gamma_{p}}}\right) + 2T_{c}K_{p} \int\limits_{\epsilon t^{\prime}}^{t^{\prime}} \frac{dy}{ \left(t^{\prime}+\frac{\tau}{2}-y\right)^{\gamma_{p}} \ln y }\right] \\
\nonumber & = & \sqrt{\ln t \ln t^{\prime}} \left[ \mathcal{O} \left(\frac{\epsilon t^{\prime}}{\left(t^{\prime}+\frac{\tau}{2}\right)^{\gamma_{p}}}\right) + \frac{2T_{c}K_{p}t^{\prime}}{\ln t^{\prime}} \times \right. \\
 & & \left. \times \int\limits_{\epsilon}^{1} \frac{du}{ \left(t^{\prime}+\frac{\tau}{2}-t^{\prime}u\right)^{\gamma_{p}} \left(1+\frac{\ln u}{\ln t^{\prime}}\right) }\right]\,,
\label{a85}
\end{eqnarray}

\noindent
where the change of variable $y=t^{\prime}u$ was performed in the last step. Since the condition $\epsilon t^{\prime}\gg 1$ is equivalent to $1 \gg -\ln\epsilon/\ln t^{\prime}=|\ln\epsilon/\ln t^{\prime}|$, one finds

\begin{eqnarray}
\nonumber C(t. t^{\prime}) \sim \sqrt{\ln t \ln t^{\prime}} \left[ \mathcal{O} \left(\frac{\epsilon t^{\prime}}{\left(t^{\prime}+\frac{\tau}{2}\right)^{\gamma_{p}}}\right) + \frac{2T_{c}K_{p}t^{\prime}}{\ln t^{\prime}} \int\limits_{\epsilon}^{1} \frac{du}{ \left(t^{\prime}+\frac{\tau}{2}-t^{\prime}u\right)^{\gamma_{p}} }\right]\,,\\
\label{a85b}
\end{eqnarray}

\noindent
which leads to

\begin{eqnarray}
C(t, t^{\prime}) \sim \frac{2^{\gamma_{p}}T_{c}K_{p}}{\gamma_{p}-1} (t^{\prime})^{1-\gamma_{p}} \left[\left(x-1\right)^{1-\gamma_{p}}-\left(x+1\right)^{1-\gamma_{p}}\right] \sqrt{1+\frac{\ln x}{\ln t^{\prime}}}\,.
\label{a86}
\end{eqnarray}

The calculation of the response function is simpler,

\begin{eqnarray}
\nonumber R(t, t^{\prime}) & \sim & f \left( \frac{\tau}{2} \right) e^{- \hat J(k_{c}) \tau} \sqrt{ 1 + \frac{\ln x}{\ln t^{\prime}} } \\
 & \sim & K_{p} 2^{\gamma_{p}} t^{\prime^{- \gamma_{p}}} \left( x - 1 \right)^{- \gamma_{p}}\sqrt{ 1 + \frac{\ln x}{\ln t^{\prime}} }.
\label{a87}
\end{eqnarray}

\indent

Therefore, the fluctuation-dissipation ratio is

\begin{eqnarray}
X(t, t^{\prime}) \sim \frac{2\left(\gamma_{p}-1\right)\ln t^{\prime}}{ 2\left(\gamma_{p}-1\right)\left[1+\left(\frac{x-1}{x+1}\right)^{\gamma_{p}}\right]\ln t^{\prime} - \left(x-1\right)\left[1-\left(\frac{x-1}{x+1}\right)^{\gamma_{p}-1}\right] }\,.
\label{a88}
\end{eqnarray}


\subsubsection*{Critical dynamics ($d>\overline{d}$)}

For $d>\overline{d}$, by equations (\ref{a67}) and (\ref{a69}), the autocorrelation is

\begin{eqnarray}
\nonumber C(t, t^{\prime}) & \sim & \mathcal{O} \left( \frac{\epsilon t^{\prime}}{ \left(t^{\prime}+\frac{\tau}{2}\right)^{\gamma_{p}} }\right) + 2T_{c} \int\limits_{\frac{\tau}{2}}^{(1-\epsilon)t^{\prime}+\tau/2} dy f(y) e^{-2\hat J(k_{c})y} \\
 & \sim & C_{eq, c}(\tau)\,,
\label{a89}
\end{eqnarray}

\noindent
since $C_{eq, c}(\tau)$ is $\mathcal{O}(1)$, which is much larger than other terms in the asymptotic limit.

\indent

The response function is obtained from equations (\ref{a39}) and (\ref{a64}):

\begin{eqnarray}
R(t, t^{\prime}) \sim f \left( \frac{\tau}{2} \right) e^{- \hat J(k_{c}) \tau}.
\label{a90}
\end{eqnarray}

\indent

The asymptotic expansions of the two-time functions in the stationary regime (for $\tau \sim 1$) are given by equations (\ref{a89}) and (\ref{a90}). In this case, one also finds that $X(t, t^{\prime}) \sim 1$. On the other hand, in the aging scenario, for $\tau\gg 1$, using (\ref{a69}), these functions have the following asymptotic behaviour:

\begin{eqnarray}
\nonumber C(t, t^{\prime}) & \sim & \mathcal{O} \left( \frac{\epsilon t^{\prime}}{ \left(t^{\prime}+\frac{\tau}{2}\right)^{\gamma_{p}} }\right) + 2T_{c}K_{p} \int\limits_{\frac{\tau}{2}}^{(1-\epsilon)t^{\prime}+\tau/2} \frac{dy}{y^{\gamma_{p}}} \\
 & \sim & \frac{2^{\gamma_{p}}T_{c} K_{p}}{\gamma_{p}-1} (t^{\prime})^{1-\gamma_{p}} \left[ \left(x-1\right)^{1-\gamma_{p}} - \left(x+1\right)^{1-\gamma_{p}} \right]
\label{a91}
\end{eqnarray}

\noindent
and

\begin{eqnarray}
R(t, t^{\prime}) \sim 2^{\gamma_{p}} K_{p} \tau^{- \gamma_{p}} = 2^{\gamma_{p}} K_{p} t^{\prime^{- \gamma_{p}}} \left( x - 1 \right)^{- \gamma_{p}}.
\label{a92}
\end{eqnarray}

\indent

In this situation, the flutuation-dissipation ratio is violated with

\begin{eqnarray}
X(t, t^{\prime}) \sim \frac{1}{1+\left(\frac{x-1}{x+1}\right)^{\gamma_{p}}}\,.
\label{a93}
\end{eqnarray}

\subsection*{Subcritical dynamics}

As in the critical dynamics, the two characteristic time scales (stationary and aging) are also present.

In the stationary case, $1 \sim \tau \ll t^{\prime}$, from equations (\ref{fasymp}), (\ref{a67}) and (\ref{a68}), one has

\begin{eqnarray}
\nonumber C(t, t^{\prime}) & \sim & M_{eq}^{4} \left[\left(t^{\prime}+\tau\right) t^{\prime}\right]^{\frac{\gamma_{p}}{2}} \left[ \frac{1}{\left(t^{\prime}+\frac{\tau}{2}\right)^{\gamma_{p}}} + \frac{2T}{\left(t^{\prime}+\frac{\tau}{2}\right)^{\gamma_{p}}} \frac{1}{2T_{c}M_{eq}^{2}} + \right. \\
\nonumber & & \left. +\frac{2T}{M_{eq}^{4}} \int\limits_{\frac{\tau}{2}}^{(1-\epsilon)t^{\prime}+\tau/2} \frac{dy f(y) e^{-2\hat J(k_{c})y}}{ \left(t^{\prime}+\frac{\tau}{2}-y\right)^{\gamma_{p}} } \right] \\
\nonumber & \sim &  \frac{ M_{eq}^{2} \left[\left(t^{\prime}+\tau\right) t^{\prime}\right]^{\frac{\gamma_{p}}{2}} }{ \left(t^{\prime}+\frac{\tau}{2}\right)^{\gamma_{p}}} \left[ M_{eq}^{2} + \frac{T}{T_{c}} + \frac{2T\left(t^{\prime}+\frac{\tau}{2}\right)^{\gamma_{p}}}{M_{eq}^{2}} \times \right. \\
\nonumber & & \left. \times \left( \int\limits_{\frac{\tau}{2}}^{\epsilon t^{\prime}+\tau/2} \frac{dy f(y) e^{-2\hat J(k_{c})y}}{ \left(t^{\prime}+\frac{\tau}{2}\right)^{\gamma_{p}} } + \int\limits_{\epsilon t^{\prime}+\tau/2}^{(1-\epsilon) t^{\prime}+\tau/2} \frac{dy f(y) e^{-2\hat J(k_{c})y}}{ \left(t^{\prime}+\frac{\tau}{2}-y\right)^{\gamma_{p}} } \right) \right] \\
\nonumber & \sim & M_{eq}^{2} \left[ 1 + \frac{2T}{M_{eq}^{2}} \frac{C_{eq, c}(\tau)}{2T_{c}} + \mathcal{O}\left( \int\limits_{\epsilon t^{\prime}+\tau/2}^{(1-\epsilon)t^{\prime}+\tau/2} \frac{\left(t^{\prime}+\frac{\tau}{2}\right)^{\gamma_{p}} dy}{ y^{\gamma_{p}}\left(t^{\prime}+\frac{\tau}{2}-y\right)^{\gamma_{p}} }\right) \right] \\
\nonumber& = & M_{eq}^{2} + \left(1-M_{eq}^{2}\right) C_{eq, c}(\tau) + \mathcal{O}\left( \left(\frac{t^{\prime}+\frac{\tau}{2}}{\epsilon t^{\prime}}\right)^{\gamma_{p}} \int\limits_{\epsilon t^{\prime}+\tau/2}^{(1-\epsilon)t^{\prime}+\tau/2} \frac{dy}{y^{\gamma_{p}}} \right)\,, \\
\label{a94}
\end{eqnarray}

\noindent
where the equation (\ref{a59}) was invoked. By choosing $\epsilon>0$ such that $1\ll (t^{\prime})^{\frac{\gamma_{p}}{2\gamma_{p}-1}}\ll\epsilon t^{\prime}\ll t^{\prime}$, one has

\begin{eqnarray}
C(t, t^{\prime}) \sim M_{eq}^{2} + \left(1-M_{eq}^{2}\right) C_{eq, c}(\tau)\,.
\label{a95}
\end{eqnarray}

Using equation (\ref{a54}), the response function, described by

\begin{eqnarray}
R(t, t^{\prime}) \sim f \left( \frac{\tau}{2} \right) \sqrt{ \frac{ f(t^{\prime}) }{ f(t) } }
\label{a97}
\end{eqnarray}

\noindent
in both time scales, behaves as

\begin{eqnarray}
R(t, t^{\prime}) \sim f \left( \frac{\tau}{2} \right) e^{- \hat J(k_{c})\tau } \left(1 + \frac{\tau}{t^{\prime}} \right)^{ \frac{\gamma_{p}}{2} } = f \left( \frac{\tau}{2} \right) e^{- \hat J(k_{c})\tau } + \mathcal{O} \left( \frac{\tau}{t^{\prime}} \right)
\label{a98}
\end{eqnarray}

\noindent
in the stationary case. It is not difficult to see that in this case the fluctuation-dissipation theorem is valid, with $X(t, t^{\prime}) \sim 1$.

In the aging regime, $1 \ll \tau \sim t^{\prime}$, from equations (\ref{fasymp}) and (\ref{a54}) in (\ref{a61}), the autocorrelation can be written as

\begin{eqnarray}
\nonumber C(t, t^{\prime}) & \sim & M_{eq}^{4} \left[\left(t^{\prime}+\tau\right)t^{\prime}\right]^{\frac{\gamma_{2}}{2}} \left[ \frac{1}{\left(t^{\prime}+\frac{\tau}{2}\right)^{\gamma_{p}}} + \frac{2T}{\left(t^{\prime}+\frac{\tau}{2}\right)^{\gamma_{p}}} \frac{1}{2T_{c}M_{eq}^{2}} + \right. \\
\nonumber & & \left. + \frac{2TK_{p}}{M_{eq}^{4}} \int\limits_{\frac{\tau}{2}}^{(1-\epsilon)t+\tau/2} \frac{dy}{ y^{\gamma_{p}}\left(t^{\prime}+\frac{\tau}{2}-y\right)^{\gamma_{p}} } \right] \\
\nonumber & = &  M_{eq}^{2} \left[\left(t^{\prime}+\tau\right)t^{\prime}\right]^{\frac{\gamma_{2}}{2}} \left[ \frac{1}{\left(t^{\prime}+\frac{\tau}{2}\right)^{\gamma_{p}}} + \mathcal{O} \left( \frac{1}{\left(\epsilon t^{\prime}\right)^{\gamma_{p}}} \int\limits_{\frac{\tau}{2}}^{(1-\epsilon)t^{\prime}+\tau/2} \frac{dy}{y^{\gamma_{p}}} \right) \right]\,.\\
\label{a99}
\end{eqnarray}

\noindent
Taking $\epsilon>0$ such that $(t^{\prime}+\tau/2)\tau^{(1-\gamma_{p})/\gamma_{p}}\ll\epsilon t^{\prime}\ll t^{\prime}$, one has

\begin{eqnarray}
C(t, t^{\prime}) \sim M_{eq}^{2} \left[ \frac{4x}{\left(x+1\right)^{2}} \right]^{\frac{\gamma_{p}}{2}}\,.
\label{a100}
\end{eqnarray}

The calculation of the response function is simpler:

\begin{eqnarray}
R(t, t^{\prime}) \sim \displaystyle\frac{ K_{p} e^{\hat J(k_{c}) \tau } }{ \left( \frac{\tau}{2} \right)^{ \gamma_{p}} } e^{- \hat J(k_{c}) \tau } x^{ \frac{\gamma_{p}}{2} } = K_{p} 2^{\gamma_{p}} t^{\prime^{- \gamma_{p}}} x^{ \frac{\gamma_{p}}{2} } \left( x - 1 \right)^{ - \gamma_{p} }.
\label{a101}
\end{eqnarray}

\indent

From equations (\ref{a100}) and (\ref{a101}), the fluctuation-dissipation ratio is given by the asymptotic expression

\begin{eqnarray}
X(t, t^{\prime}) \sim \frac{2TK_{p}}{\gamma_{p}M_{eq}^{2}} t^{\prime^{1 - \gamma_{p}}} \left( \frac{x + 1}{x - 1} \right)^{1 + \gamma_{p}}.
\label{a102}
\end{eqnarray}

\subsection*{Technical note}

From equations (\ref{a61}) and (\ref{volterra}), with $t\rightarrow t^{\prime }+\tau /2$, one finds another form for the autocorrelation,

\begin{eqnarray}
\nonumber C(t, t^{\prime}) & = & \frac{1}{\sqrt{\psi(t)\psi(t^{\prime})}} \left[ \psi\left(t^{\prime}+\frac{\tau}{2}\right) - 2T \int\limits_{t^{\prime}}^{t^{\prime}+\frac{\tau}{2}} dy f\left(t^{\prime}+\frac{\tau}{2}-y\right)\psi(y)\right] \\
 & = & \frac{1}{\sqrt{\psi(t)\psi(t^{\prime})}} \left[ \psi\left(t^{\prime}+\frac{\tau}{2}\right) - 2T \int\limits_{0}^{\frac{\tau}{2}} dy f(y)\psi\left(t^{\prime}+\frac{\tau}{2}-y\right)\right]\,.
\label{a103}
\end{eqnarray}

It is possible to use this expression for calculating the asymptotic forms of autocorrelation. One then recovers the same asymptotic results that have already been reported, with the exception a discrepancy in the critical dynamics for $d>\overline{d}$, which will not display any aging behaviour. Equation (\ref{a103}), however, involves strongly varying terms, which may even change sign, and whose asymptotic behavior may turn out to be much more difficult to analyse. 


\end{document}